\documentclass[a4paper,12pt,twoside,english]{article}
\usepackage[latin1]{inputenc}
\usepackage{parskip}               
\usepackage[dvips]{graphicx,epsfig,color}
\usepackage{wrapfig,rotating}
\usepackage{amsmath,wasysym,array}
\usepackage[thinspace, thickqspace, amssymb, Gray]{SIunits}
\usepackage{hhline}
\usepackage{lineno}
\linenumbers
\usepackage{cite}                  
\usepackage[ , ,bf,it]{caption}    

\newlength{\dinwidth}
\newlength{\dinmargin}
\newcommand{\artitle}[1]{}
\newcommand{\artitlekeep}[1]{{\em``#1''}}
\newcommand{\geant}         {{\sc Geant4}}
\newcommand{\goodfellow}    {{\sc GoodFellow}}
\newcommand{\hamamatsu}     {{\sc Hamamatsu}}
\newcommand{\perkinelmer}   {{\sc PerkinElmer}}
\newcommand{\photonis}      {{\sc Photonis}}
\newcommand{\CfFt}          {\ensuremath{ \mbox{C}_4\mbox{F}_{10} }}
\newcommand{\corrto}        {\ensuremath{ \hat{=} }}

\definecolor{dblack}   {rgb}{0.00, 0.00, 0.00}  \newcommand{\dblack}     {\color{dblack}}

\begin{document}

\begin{titlepage}
  \begin{flushleft}
    {\tt DESY 10-225    \hfill    ISSN 0418-9833} \\
    {\tt November 2010, update Juli 2011}                  \\
  \end{flushleft}

  \vspace{1.0cm}
  \begin{center}
    \begin{Large}
      {\bfseries \boldmath Design and Construction of a Cherenkov Detector 
        for Compton Polarimetry at the ILC}

      \vspace{1.5cm}
      Christoph Bartels$^{1,2}$, Joachim Ebert$^2$, Anthony Hartin$^1$, \\
      Christian Helebrant$^{1}$, Daniela K\"afer$^1$, and Jenny List$^1$
    \end{Large}

    \vspace{.3cm}
    1- Deutsches Elektronen-Synchrotron DESY \\ 
    Notkestr. 85,  22607 Hamburg, Germany
    \vspace{.1cm} \\
    2- Universit\"at Hamburg, Institut f\"ur Experimentalphysik \\
    Luruper Chaussee 149,  22761 Hamburg, Germany
  \end{center}

  \vspace{1cm}

  \begin{abstract}
    This paper describes the design and construction of a Cherenkov detector conceived 
    with regard to high energy Compton polarimeters for the International Linear Collider, 
    where beam diagnostic systems of unprecedented precision must complement the 
    interaction region detectors to pursue an ambitious physics programme. 
    Besides the design of a prototype Cherenkov detector, detailed simulation studies are
    presented. Results of a first testbeam campaign with the main objective of validating the
    simulation in terms of the light distribution inside the channels and the channel response 
    are presented. Furthermore, a new method for aligning the detector without the need of
    dedicated data taking has been developped.
  \end{abstract}

  \vspace{1.0cm}
  \begin{center}
    Submitted to JINST
  \end{center}

\end{titlepage}

\section{Introduction}
The measurement and control of beam parameters to permille level precision will play 
an important role in the physics programme~\cite{bib:RDR, bib:POWER} of the 
International Linear Collider (ILC). 
For electroweak processes, the absolute normalisation of expected event rates depends 
on both, luminosity and polarisation. The luminosity will be measured to a precision 
of $10^{-3}$ to $10^{-4}$, while for the luminosity weighted polarisation average an 
accuracy of $10^{-3}$ seems achievable~\cite{bib:Marchesini-PhD}.

While for beam energy and luminosity measurements the ILC's precision goals have already 
been achieved at previous colliders, polarimetry has to be improved by at least a factor 
of two compared to the most precise previous measurement of the SLD polarimeter~\cite{bib:SLDpolmeas}. 

The polarisation determination at the ILC will combine the measurements of two de\-dicated 
Compton polarimeters, located upstream and downstream of the $e^+e^-$ interaction point, 
with measurements from the $e^+e^-$ interactions themselves. While the $e^+e^-$ data 
will finally yield the absolute polarisation scale, the polarimeters provide fast measurements 
which allow to track variations over time and to detect possible correlations with the 
luminosity or the polarisation of the other beam. Therefore, each polarimeter has to 
reach a systematic accuracy of at least $\delta{\mathcal P}/{\mathcal P}=0.25$\%\footnote{for typical 
ILC beam polarisation values of ${\mathcal P}_{e-} \geq 80$\% and 
  $\mathcal{P}_{e+} \geq 30\%$ or even $\geq 60\%$}. 
Two polarimeters per beam are required in order to measure the polarisation of the beams 
in collisions. Both polarimeters have been designed for operation at beam energies between 
$45\:\giga\electronvolt$ and $500\:\giga\electronvolt$. A detailed description of the polarimeters 
can be found in~\cite{bib:EP-paper}. 

Both polarimeters make use of the polarisation dependence of Compton scattering to ensure 
a non-destructive measurement of the longitudinal beam polarisation. Circularly polarised 
laser light is shot under a small angle onto the individual bunches causing typically in the 
order of $10^3$ electrons\footnote{or positrons in case of the positron beam of 
  the ILC which is equipped analogously.} per bunch to undergo Compton scattering.
The energy spectrum of these scattered particles depends on the product of laser and beam 
polarisations, so that the differential rate asymmetry with respect to the laser helicity 
is directly proportional to the beam polarisation. 
Since the scattering angle in the laboratory frame is less than 10~$\micro$rad, a magnetic 
chicane is used to transform the energy spectrum into a spatial distribution which is 
then measured by Cherenkov detectors.

\noindent A Cherenkov detector was chosen for several reasons: 
\begin{enumerate}
  \item[(i)] In combination with the magnetic chicane, it allows to measure 
    the energy  spectrum of many electrons arriving simultaneously. With about 
    $10^3$ Compton interactions per electron bunch, a statistical precision of 
    $\delta\mathcal{P} = 1\%$ is achieved for each of the about 
    3000 bunch positions in a train after only 
    20 trains ($\corrto\; 4$ seconds). 
    For the average polarisation of all bunch positions, this corresponds to a 
    statistical error below $0.1\%$ after $1$ second~\cite{bib:EP-paper}.
  \item[(ii)] For relativistic electrons ($\beta = v/c \approx 1$), 
    Cherenkov radiation is independent of the electron energy.  Thus, the 
    number of Cherenkov photons will be directly proportional to the number 
    of electrons per detector channel.
  \item[(iii)] Typical Cherenkov media like gases or quartz are sufficiently 
    radiation hard to withstand the flux of $10^7$ electrons passing through 
    the detector per second. 
\end{enumerate}

Developing a Cherenkov detector suitable for achieving the target precision of 
$\delta{\mathcal P}/{\mathcal P}=0.25$\%
demands improvements in various areas of the experimental setup. On the detector side, 
especially the linearity of the detector response, its homogeneity with respect to the entrance 
point of the Compton electrons as well as the detector's alignment to the beam axis are important 
to control.
In order to test these and further aspects, a prototype detector has been 
designed, simulated and constructed as described in the chapters~\ref{sec:simulation} 
and~\ref{sec:construction}. The prototype has been operated in a testbeam 
campaign at the ELSA stretcher ring in Bonn, with the main objective of validating the 
simulation in terms of the channel acceptance and light distribution inside the channels 
employing multi-anode photomultipliers. The results of these studies are presented in chapter~\ref{sec:testbeam}, including a new method for aligning the detector directly 
from polarimetry data without the need of expensive dedicated alignment data taking.

\section{Detector Design and Simulation}  \label{sec:simulation}
\begin{figure}[!h]
  \begin{picture}(16.0, 11.0)
    \put( 0.00, 0.60)  {\epsfig{file=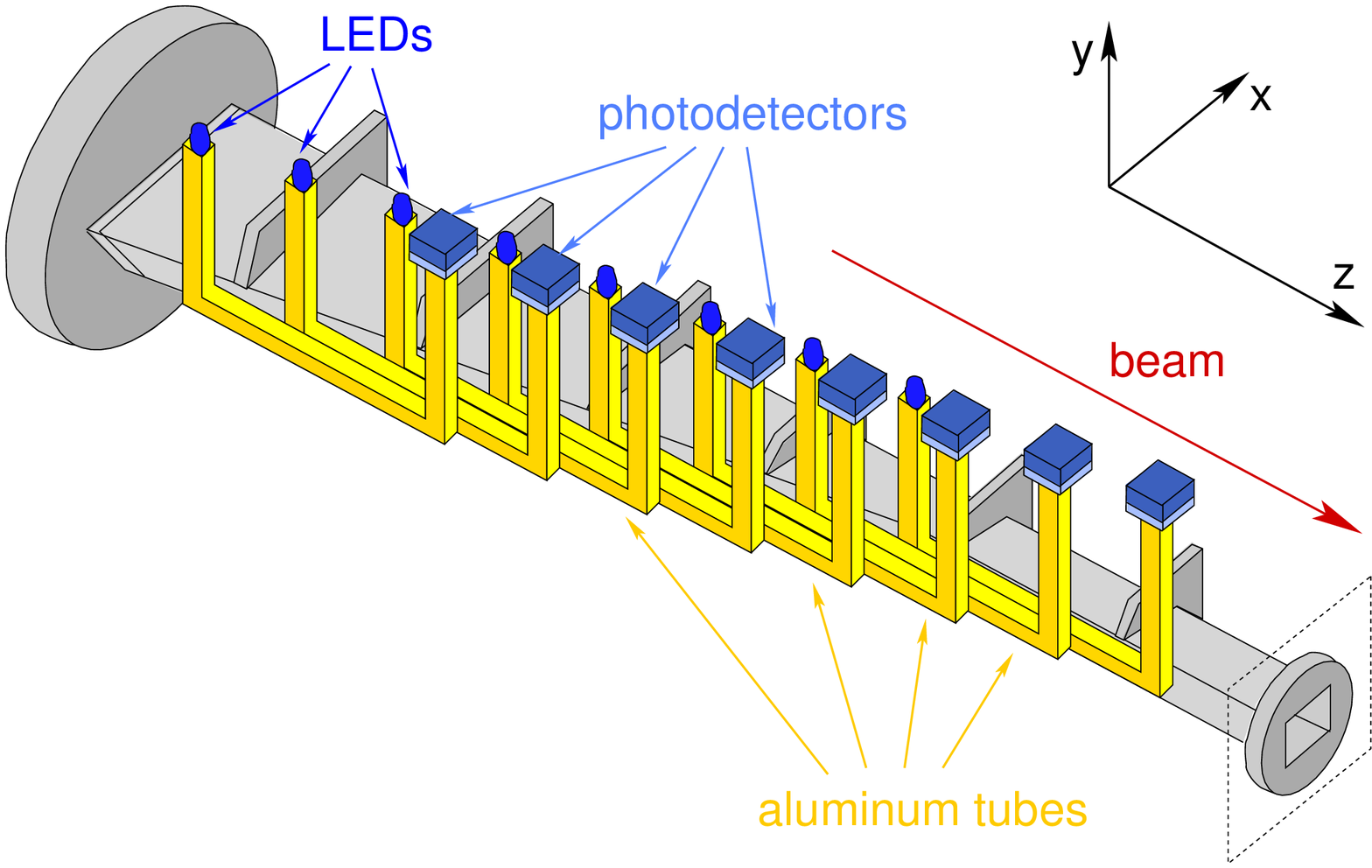, clip= , width=1.00\linewidth}}
    \put(-0.10, 0.00)  {\epsfig{file=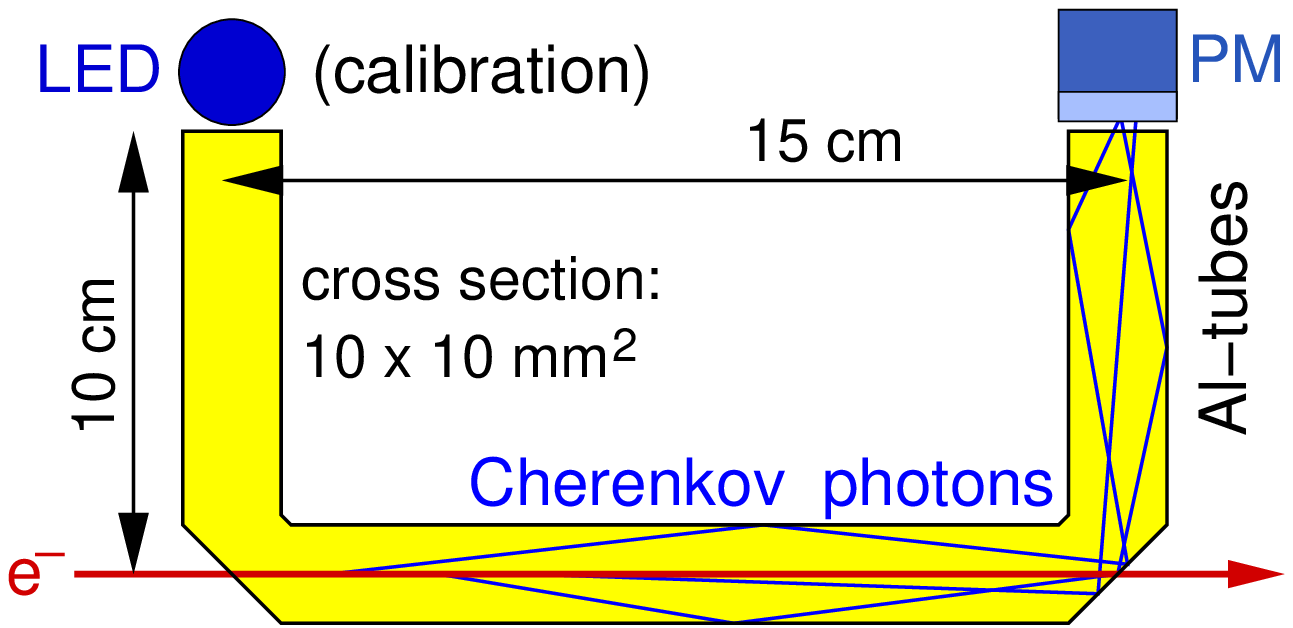,    clip= , width=0.5\linewidth}}
    \put( 0.00,10.40)  {(a)}
    \put( 0.00, 4.40)  {(b)}
  \end{picture}
  \caption{\label{fig:ILDpol-CherTube} \it 
    (a) Illustration of a Cherenkov detector for ILC polarimetry, 
    here for better visibility with eight instead of the actually 
    forseen 20 readout channels; and 
    (b) sketch of one such gas-filled aluminium channel.
  }
\end{figure}

A conceptual design for the Cherenkov detector envisioned for the ILC polarimeters is shown 
in Figure~\ref{fig:ILDpol-CherTube}(a). It will consist of staggered `U-shaped' aluminium 
channels lining the tapered exit window of the beam pipe, similar to the design proposed in~\cite{bib:TESLA-note}. The channels are filled with a Cherenkov gas 
so that relativistic electrons traversing their base emit Cherenkov radiation which is 
reflected upwards in the hind U-leg to the photodetectors. A single channel is sketched 
in Figure~\ref{fig:ILDpol-CherTube}(b).

\noindent The wavelength spectrum of Cherenkov radiation is given by: 

    \begin{equation}
      \frac{dN^{\gamma}}{d\lambda} = 2\pi\alpha \, \left(1-\frac{1}{n^2\beta^2}\right) \frac{1}{\lambda^2}\;\ell\,, 
      \hspace*{8mm}\mbox{with:} \hspace*{4mm}
      \left.
        \begin{array}{ll}
          N^{\gamma} &: \mbox{ mean number of photons,}  \\
          \lambda    &: \mbox{ wavelength,} \\
          \alpha     &: \mbox{ fine structure constant,} \\
          n          &: \mbox{ radiator's refractive index,} \\
          \beta      &: \mbox{ velocity } (\beta=\frac{v}{c}), \\
          \ell       &: \mbox{ radiator length}
        \end{array}
      \right.
      \label{eq:CherRad}
    \end{equation}

While the velocity ($\beta$) can be regarded as constant for electron energies relevant 
at the ILC, the refractive index depends on the wavelength, as well as on the temperature 
and the gas pressure. At small wavelengths, the refractive index typically rises like 
$(n-1) = A / [ \lambda_0^{-2} - \lambda^{-2}]$. This behaviour has been measured for \CfFt{} 
for the Ring Imaging Cherenkov Detector of the DELPHI experiment at LEP~\cite{bib:DELPHI-RICH1}. 
Furthermore, $n-1$ is proportional to the number density of molecules for $n \approx 1$~\cite{bib:maxwells-formula} 
and thus increases proportionally with the inverse temperature and the pressure.

To simplify further references, a right-handed coordinate system, as shown in 
Figure~\ref{fig:ILDpol-CherTube}, will be used throughout the rest of this publication. 
Assuming the electron beam travels in positive $z$-direction, the $y$-axis points upwards, 
and the $x$-axis to the left when looking in the direction of the electron beam.

\subsection{Requirements and conceptual design} \label{sub:sim-concept}
The design of the Cherenkov detectors for the ILC polarimeters is driven by the
requirements listed in the following. In most cases, the exact values will depend on the
final design of the accelerator, of the magnetic chicane and of the laser system, which is still
subject to change. Therefore only indicative numbers are given here.
\begin{description}
  \item[Dynamic Range] ~~ \\ 
    The detector has to be able to cover a dynamic range from $\mathcal{O}(1)$ to several 
    hundreds of Compton electrons per channel and bunch crossing. This applies especially
    to the channels near the Compton edge, where the asymmetry with respect to the laser 
    helicity is largest~\cite{bib:EP-paper}.
  \item[Homogeneous detector response] ~~ \\
    At the SLD polarimeter, variations of the detector response   
    with respect to the electrons' entry positions lead to a correction on the analyzing 
    power in the order of $1\%\pm0.5\%$~\cite{bib:SLDEliaKing}. Although part of this 
    effect was due to a preradiator, which is not forseen to be employed at the ILC, the SLD
    example shows that the homogeneity of the detector response needs to be considered and
    should not contribute more that a permille to the total uncertainty budget.  
  \item[Alignment] ~~ \\
    The Compton edge position has to be controlled to $\mathcal{O}(100~\mu\mathrm{m})$ in order
    to keep the impact on the analyzing power below $0.1\%$. Tilts of the detector typically result
    in changes of the analyzing power in the order of 0.05\%/mrad, depending on the considered channel 
    and rotation axis. These numbers are similar to those observed at the SLD  
    polarimeter~\cite{bib:SLDEliaKing}. At SLD the alignment was monitored regularly in a dedicated 
    operation mode by moving the detector with respect to the beam and interpolated in between
    these calibration runs. While this approach could be followed at the ILC, it costs expensive
    accelerator operation time for which no polarisation measurement can be provided. Therefore 
    the development and test of alignment methods which can be performed during polarimetry
    data taking are amoung the goals of the prototype presented here.   
  \item[Robustness with respect to backgrounds] ~~ \\
    A gas with a Cherenkov threshold in the \mega\electronvolt-regime should be used to avoid 
    the emission of Cherenkov light from low energetic electrons and muons, e.g. from the beam halo, 
    from beam-gas interactions, or electrons pair-produced from synchrotron radiation.
    A layout allowing the photodetectors and electronics to be placed well outside the 
    beam-plane is mandatory. 
  \item[Detector Linearity \& Calibration system] ~~ \\
    The linearity of the detector response has to be controlled to a level of about $0.75$\%
    in order to limit the effect on the analyzing power to $0.1$\%. Therefore,
    a dedicated calibration system is forseen to monitor the detector response, 
    especially its linearity, in-situ. This could be realised by equipping the channels 
    with light emitting diodes (LEDs) which should also be placed outside the beam-plane.
    Such a system could collect data during breaks in the accelerator operation, or even 
    in-between two ILC bunch trains.
\end{description}

The chosen channel geometry is illustrated in Figure~\ref{fig:ILDpol-CherTube}. The (Compton scattered) electrons traverse the horizontal U-base, while the upward pointing legs serve to place the 
photodetectors and the calibration 
system outside of the beam-plane. With increasing length of the U-base more Cherekov light 
is produced, but on the other hand the alignment requirements become more stringent and 
additional reflections will decrease the light yield. Simulations suggest that a length 
of $15\:\centi\meter$ is a reasonable choice.

 In contrast to the ILC-like design of 20 staggered channels 
(c.f.\ Figure~\ref{fig:ILDpol-CherTube}(a)), the prototype detector consists of two parallel, 
non-staggered channels. Apart from this difference, the 
prototype detector allows to test all relevant aspects of the full detector.  Especially 
it will serve in the future as a test bed for the calibration system which is currently 
under development as well as for the final choice of photodetectors. 

\subsection{Optical simulation} \label{sub:sim-geant}
A detailed simulation of the prototype detector based on \geant~\cite{bib:geant4} has been 
created in order to support the design process and the interpretation of the testbeam data. 

For electrons and positrons Cherenkov radiation, multiple scattering, ionisation, brems\-strahlung 
and annihilation are simulated. Apart from annihilation, the same processes are taken into account 
for muons, which are relevant when studying the impact of accelerator background. 
For the Cherenkov photons optical processes have to be considered since their wavelengths are much 
larger than a typical atomic spacing. In particular, absorption in the photodetector 
entrance window, boundary effects (like reflection and absorption) at the channel walls, as well 
as Rayleigh elastic scattering have been included in the simulation~\cite{bib:geant4_physics}. 

As Cherenkov gas, perfluorobutane (\CfFt) has been chosen due to its high threshold of $10\:\mega\electronvolt$, 
which makes the detector robust against background from low energetic charged particles. 
The wavelength dependence of the refractive index is implemented in the simulation according
to~\cite{bib:DELPHI-RICH1}. Since the polarisation measurement is based on rate asymmetries, 
it will be insensitive to the exact value of $n$. Also variations of $n$ with time which are 
slow with respect to the laser helicity flipping rate (like thermal variations) 
will cancel out within the asymmetry.  
Therefore such effects are currently not simulated and the temperature 
and gas pressure inside the detector box are set to $T=20\:\celsius\,$ 
and $\,p=1\:\mathrm{atm} = 1.01325\;\mathrm{bar}$, respectively.

Pure \CfFt{} is fully transparent even in the far UV range. In the presence of impurities, 
especially water or oxygen, the transparency can drop significantly for wavelengths smaller 
than $200\:\nano\meter$~\cite{bib:DELPHI-RICH2}, where two of the four employed photodetectors 
are sensitive, c.f.\ Section~\ref{sub:const-photodet}.  Since the precise knowledge of the 
absolute photon yield is not crucial for the rate asymmetry measurement, gas impurities have 
not been implemented in the simulation.

Two different types of aluminium have been implemented in the simulation according to the 
reflectivity studies summarised in Section~\ref{sub:sim-reflect}. Three of the four walls of each channel are made of diamond-milled aluminium, while the inter-channel wall consists of two $150\:\micro\meter$ sheets of rolled aluminium. 
The wavelength dependency is interpolated linearly between the values listed in Table~\ref{tab:reflect-meas}. 
While for the diamond-milled walls the absolute normalisation is fixed to the values in Table~\ref{tab:reflect-meas}, the reflectivity of the rolled sheets can be adjusted via the ratio of the two materials' reflectivities $r := R_{\rm roll}^{\rm eff}/R_{\rm diam}$, which is a parameter of the simulation.

\begin{figure}[!h]
  \begin{minipage}[c]{0.38\linewidth}
    \caption{\label{fig:SimGEANT} \it 
      Event display of the 2-channel
      prototype simulation: \newline
      The electron beam (red) passes from left to right through the U-base of 
      the aluminium tubes filled with perfluorobutane, \CfFt, and emits Cherenkov 
      photons (green). These are reflected upwards to a photodetector mounted 
      on the hind U-leg. 
      The channels are separated by a thin foil (light grey). \newline
      Due to a surrounding gas-filled box (not shown), Cherenkov radiation can 
      also be emitted before/after the electron beam enters/exits the aluminium 
      tubes, but it cannot reach the photodetector.
    }
  \end{minipage}\hspace*{2.5mm}
  \begin{minipage}[c]{0.62\linewidth}
    \begin{picture}(8.0, 8.5)
      \put( 0.10, 0.00)  {\epsfig{file=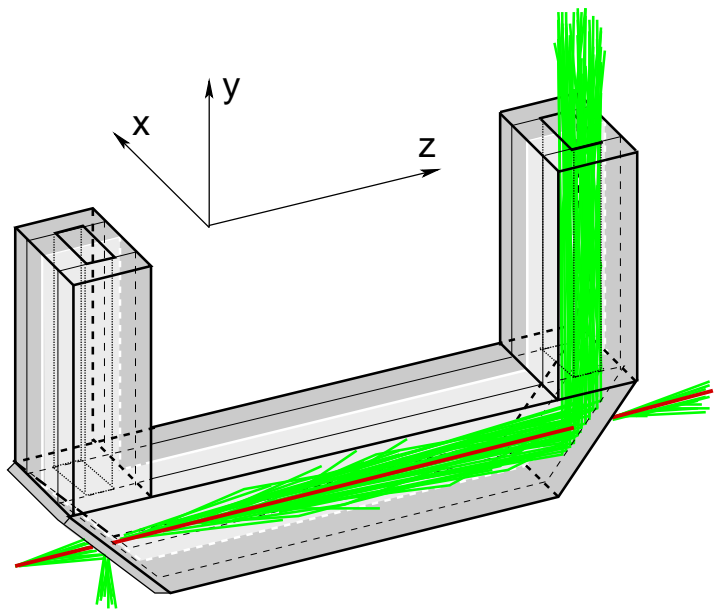, clip= , width=0.96\linewidth}}
    \end{picture}
  \end{minipage}
\end{figure}

Figure~\ref{fig:SimGEANT} shows the channel structure with a single electron (red line) 
passing from left to right through the U-base of the right-hand side channel. 
It emits Cherenkov light (green), which is reflected upwards at the end of the U-base 
towards the photodetector.  Cherenkov light produced outside the channel structure in 
the ambient gas cannot reach the photodetectors. 
The optical simulation ends at the photocathode and all Cherenkov spectra are stored 
for further processing and digitisation.

Unless specified otherwise, the simulations shown in the following sections where performed in view of the testbeam situation at ELSA with a two-dimensional Gaussian with standard deviations $\sigma_{x}=\sigma_{y}=0.5\:\milli\meter$ as beam profile and zero divergence over the length of the detector prototype. Per beam position, $10^{6}$ electrons with an energy of $2\:\giga\electronvolt$ have been simulated.

\subsection{Cherenkov spectra and light distributions} \label{sub:sim-lightdist}
The number of emitted Cherenkov photons follows a Poisson distribution with the mean value 
$N^{\gamma}$ given by the integral of Formula~\ref{eq:CherRad} over the relevant wavelength range. The number of photons reaching the photocathode is expected to follow a Poisson distribution as well, as long as other contributions, for instance from electrons showering in the channel walls or multiple-scattering in the entrance window, are small.

To study the influence of the detector geometry and the reduced reflectivity of the inter-channel 
wall on the expected light distribution, the simulation has been run twice with  different reflectivities for the inter-channel wall and the beam centered on one of the two channels. 

\begin{figure}[!h]
  \begin{picture}(16.0, 8.0)
    \put(-0.05, 0.00)  {\epsfig{file=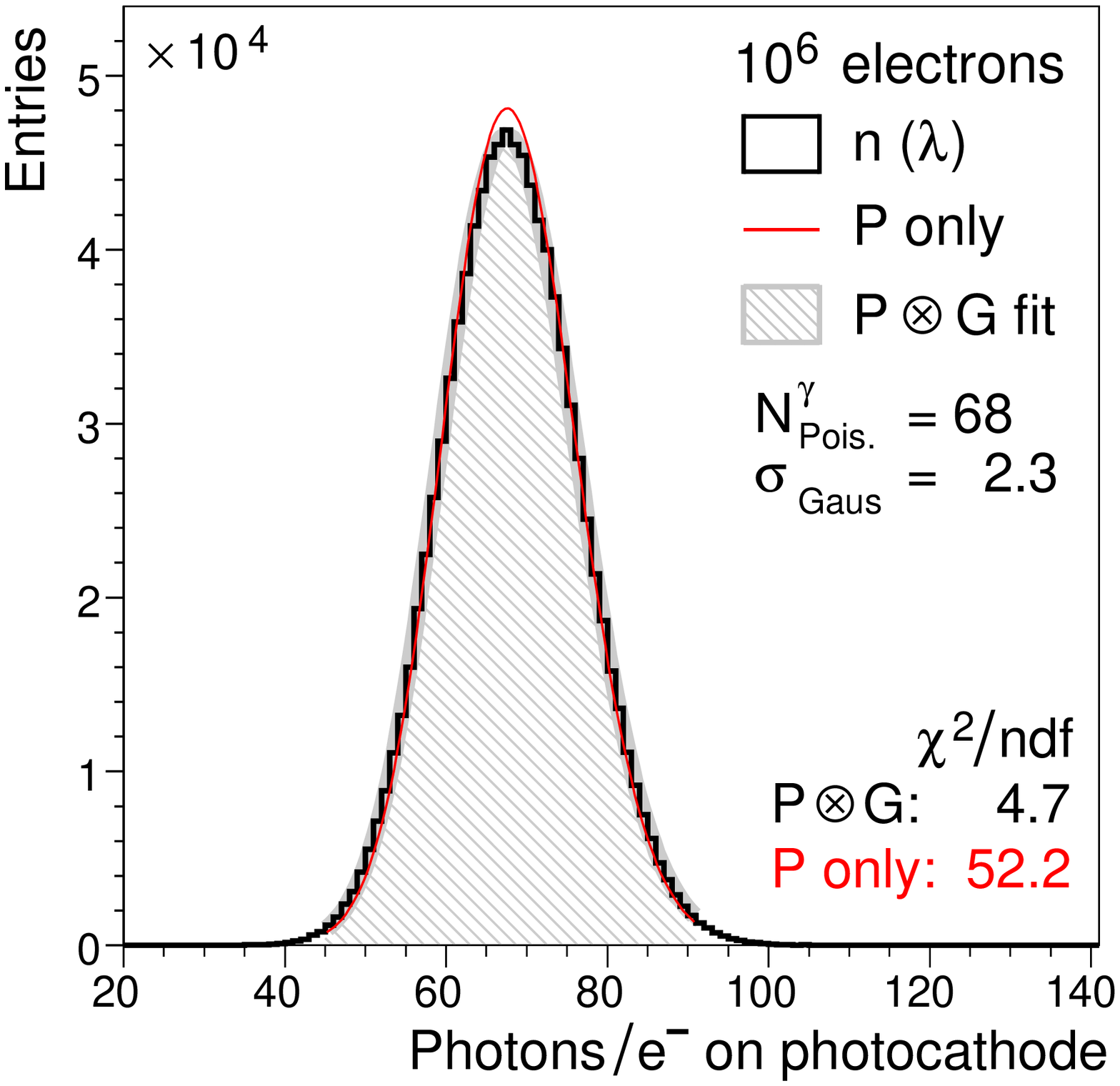, clip= , width=0.50\linewidth}}
    \put( 8.15, 0.00)  {\epsfig{file=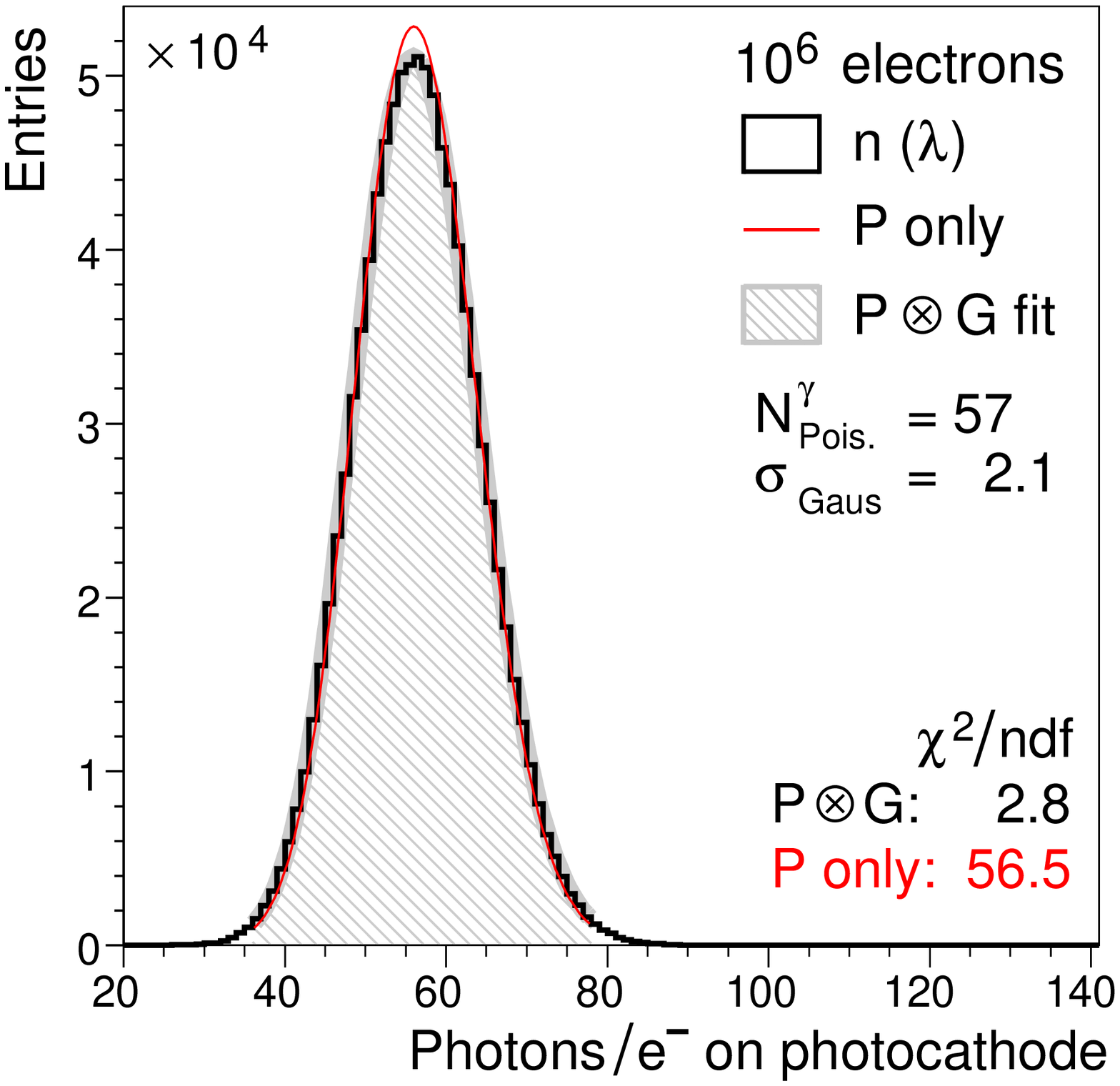, clip= , width=0.50\linewidth}}
    \put( 0.00, 0.00)  {(a)}
    \put( 8.35, 0.00)  {(b)}
  \end{picture}
  \caption{\label{fig:gammaperelectron-sim} \it 
    Average number of photons reaching the photocathode 
    per $2\:\giga\electronvolt$ electron: \newline
    (a) with equal reflectivities for all channel walls and 
    (b) with a reduced reflectivity of the inter-channel wall 
    ($r = 50\%$).
  }  
\end{figure}

Figure~\ref{fig:gammaperelectron-sim} shows the number of Cherenkov photons reaching the photocathode for both cases: In Figure~\ref{fig:gammaperelectron-sim}(a), all channel walls have been simulated 
with the reflectivity of diamond-milled aluminium, resulting in on average $N^{\gamma} = 68$~photons per $2\;\giga\electronvolt$ electron. When the reflectivity of the inter-channel wall is reduced to  $r = 50\%$, about $16\%$ less photons reach the photocathode, so that the average decreases to  $N^{\gamma} = 57$~photons as shown in Figure~\ref{fig:gammaperelectron-sim}(b). 

In both cases, the distributions of the number of photons reaching the photocathode is not exactly Poissonian, but exhibit a small Gaussian broadening {\dblack due to multiple scattering in the entrance window}. For illustration, both a pure Poissonian (labeled $P$) as well as a Poisson convoluted with a Gaussian (labeled $P \otimes G$) have been fitted to the distributions. The respective $\chi^2$ per degree of freedom values clearly prefer the convoluted fit over the pure Poissonian. But with $\sigma_{\mathrm{Gauss}} \approx 2$ photons, the broadening is small compared to the mean number of photons for both reflectivity scenarios and is expected to be negligible for the much higher electron energies relevant in the polarisation measurement.

\begin{figure}[!h]
  \begin{picture}(16.0, 8.0)
    \put(-0.05, 0.00)  {\epsfig{file=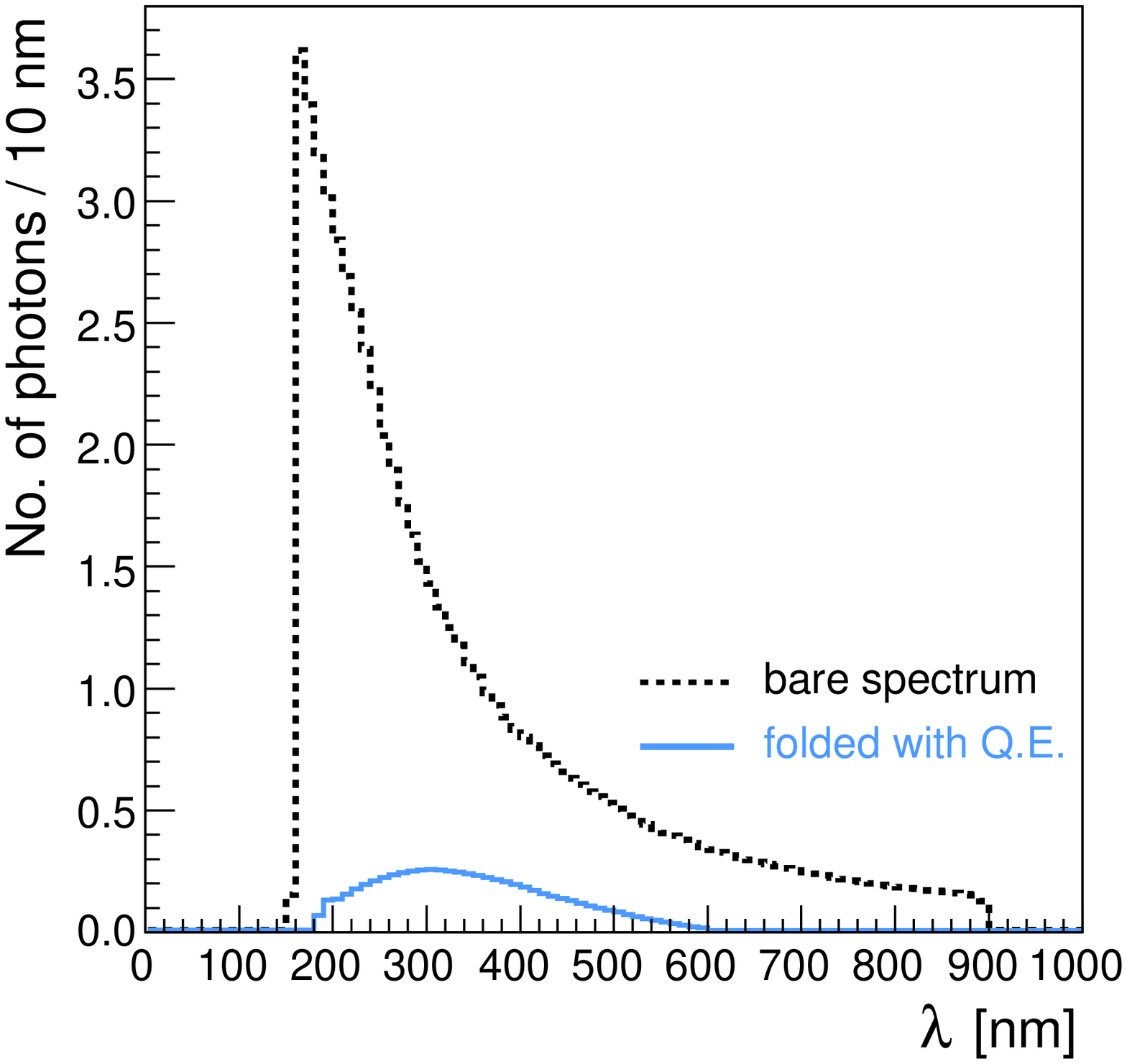, clip= , width=0.50\linewidth}}
    \put( 3.60, 3.35)  {\epsfig{file=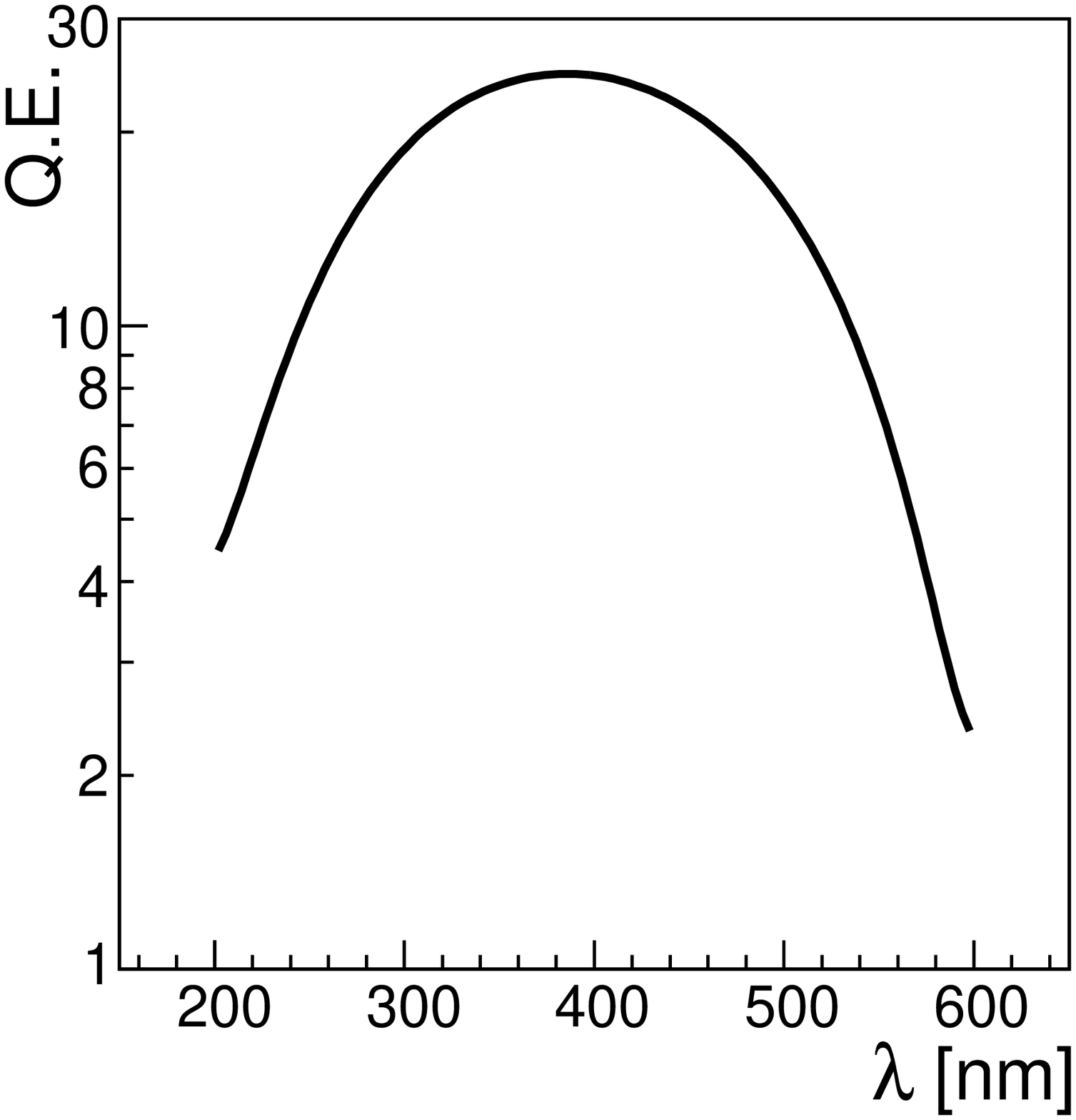,                           clip= , width=0.24\linewidth}}
    \put( 8.15, 0.00)  {\epsfig{file=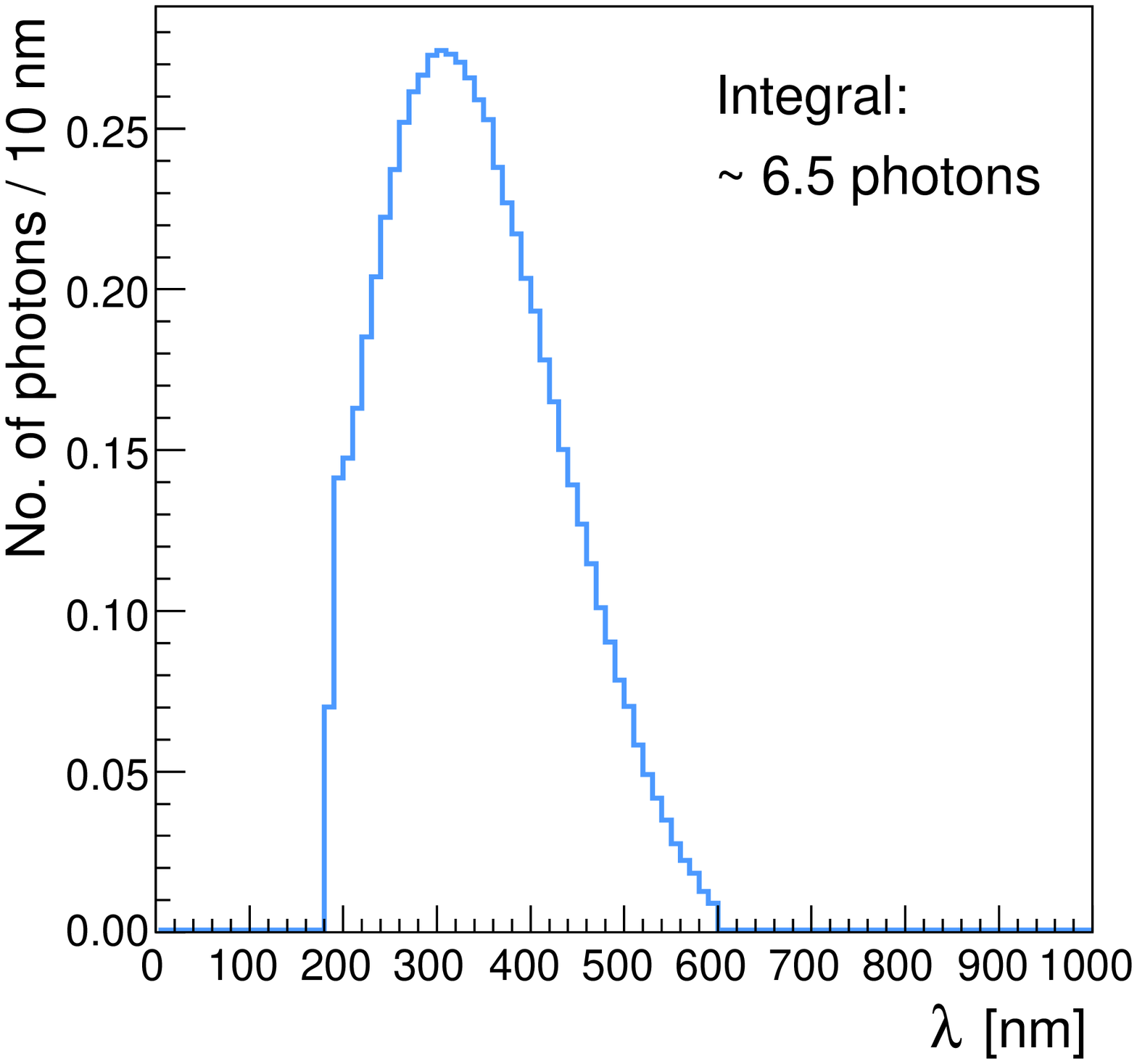,  clip= , width=0.50\linewidth}}
    \put( 0.00, 0.00)  {(a)}
    \put( 8.35, 0.00)  {(b)}
  \end{picture}
  \caption{\label{fig:CherSpec-Sim+QE} \it 
    Cherenkov spectra: 
    (a) at the photocathode (dotted line) and 
    (b) convoluted with the quantum efficiency (Q.E.) of the $2\!\times\!2$ MAPM (R7600U-03-M4), see the insert in (a). 
    The convoluted spectrum is also superimposed in (a) as the solid line.
  }
\end{figure}

Figure~\ref{fig:CherSpec-Sim+QE} shows the wavelength spectrum of these photons obtained from 
the simulation with the reduced reflectivity for the inter-channel wall. 
Figure~\ref{fig:CherSpec-Sim+QE}(a) illustrates the distribution at the photocathode, showing 
the expected $1/\lambda^2$~wavelength dependence for Cherenkov radiation. It is cut off 
below~$\lambda_{\mathrm{low}}  = 160\:\nano\meter$  and 
above~$\lambda_{\mathrm{high}} = 900\:\nano\meter$  in the simulation 
since the photodetectors are not sensitive outside this wavelength range.
Figure~\ref{fig:CherSpec-Sim+QE}(b) shows the wavelength spectrum after convolution with the 
quantum efficiency of the $2\!\times\!2$ multi-anode photodetector (\hamamatsu~R7600U-03-M4). 
This quantum efficiency is shown in the insert in Figure~\ref{fig:CherSpec-Sim+QE}(a)~\cite{bib:Hamamatsu}. 
On average, $6.5$~photons are detected corresponding to the integral of the histogram 
in Figure~\ref{fig:CherSpec-Sim+QE}(b).

\begin{figure}[!h]
  \begin{picture}(16.0, 7.8)
    \put(-0.25, 0.00)  {\epsfig{file=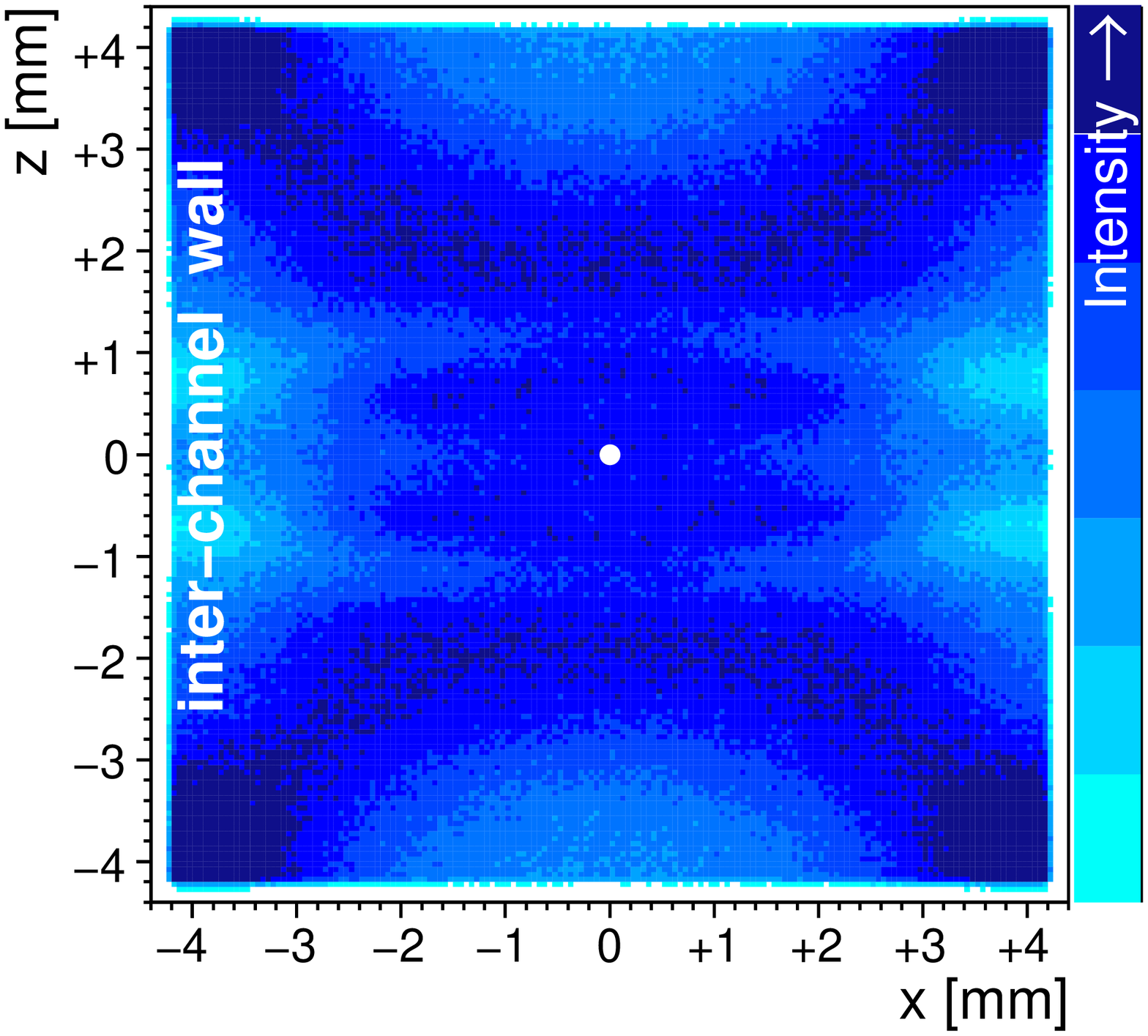, clip= , width=0.50\linewidth}}
    \put( 8.00, 0.00)  {\epsfig{file=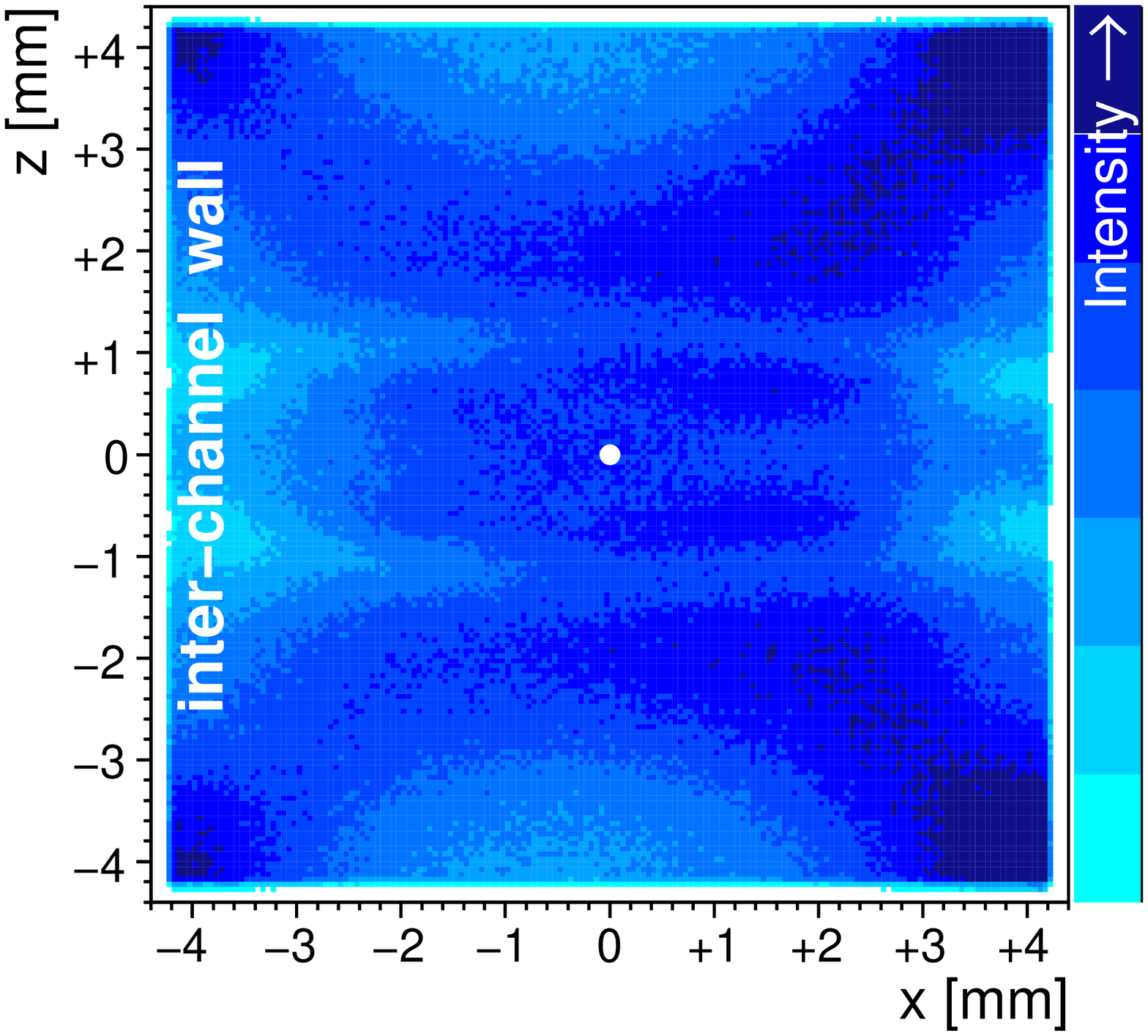, clip= , width=0.50\linewidth}}
    \put( 0.00, 0.00)  {(a)}
    \put( 8.20, 0.00)  {(b)}
  \end{picture}
  \caption{\label{fig:LightYield-comp} \it 
    Light distribution on the photocathode: 
    (a) with equal reflectivities for all channel walls, 
    (b) with a reduced reflectivity for the inter-channel wall 
    at $x=-4.25\:\milli\meter$. 
    The white dot indicates the channel centre; the intensity scale
     ranges from 40\% to 100\%. 
  }
\end{figure}

Figure~\ref{fig:LightYield-comp} shows the resulting spatial light distributions on the photocathode, again for two different assumptions on the reflectivity of the inter-channel wall.
The result obtained with equal reflectivities for all inner surfaces is illustrated in 
Figure~\ref{fig:LightYield-comp}(a).  In this case, the observed non-uniformities are due to 
the influence of the detector geometry.  For reference a white dot indicates the position of 
the channel centre.  The distribution exhibits an X-like structure of increased photon yield 
which is symmetrical about the~$x$- and $z$-axes.  A reduced light yield is visible in two 
narrow bands at $z=\pm 1\;\milli\meter$ as residuals from the $90^\circ$ reflection at the 
end of the Cherenkov section. 
Lowering the reflectivity of the inter-channel wall (located at $x=-4.25\;\milli\meter$) 
changes the symmetry of the light pattern as shown in Figure~\ref{fig:LightYield-comp}(b). 
Near the inter-channel wall, the light yield is reduced, but the X-like structure remains.

\begin{figure}[!h]
  \begin{picture}(16.0, 8.3)
    \put(-0.05, 4.80)  {\epsfig{file=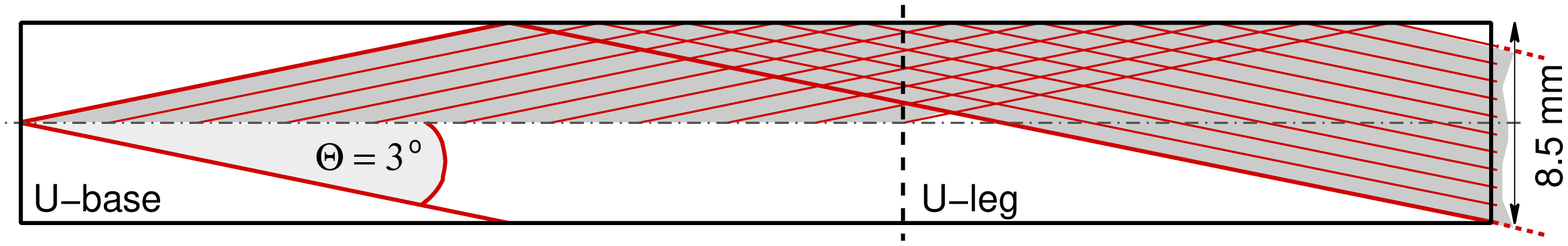,       clip= , width=1.01\linewidth}}
    \put(-0.05, 0.00)  {\epsfig{file=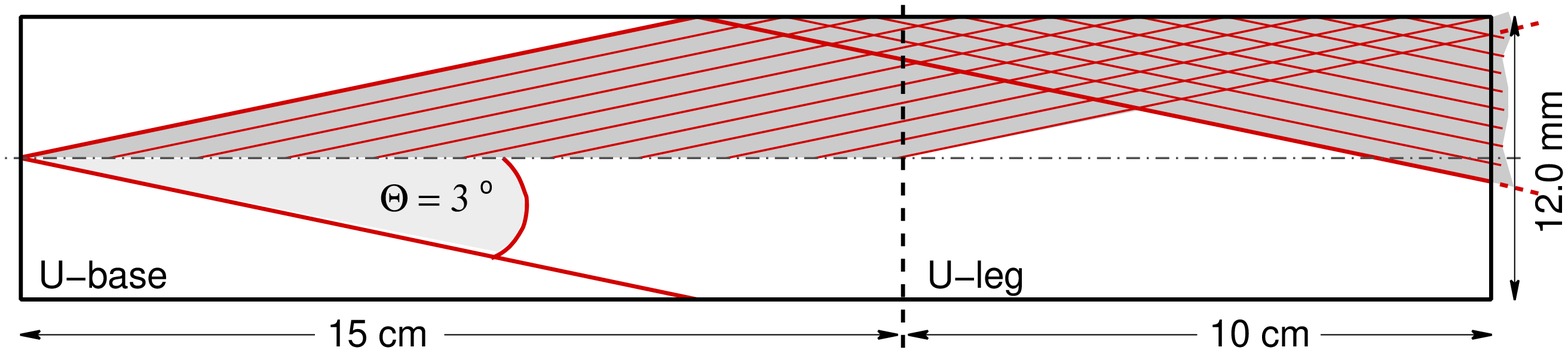, clip= , width=1.01\linewidth}}
    \put( 0.00, 7.45)  {(a) photons emitted in the horizontal/vertical plane}
    \put( 0.00, 3.95)  {(b) photons emitted toward the channel corners}
  \end{picture}
  \caption{\label{fig:reflection-sketch} \it 
    Sketches of possible light paths for electrons traversing the channel along the central axis 
    and the Cherenkov angle of the chosen gas $(\Theta_{\rm Ch} = 3^{\circ})$.  The channel aspect ratio 
    has been enlarged by a factor of 4 for better visibility. \newline
    (a) Photons emitted in the horizontal/vertical plane illuminate the entire channel width at the photocathode, while 
    (b) photons emitted towards the channel corners only illuminate half the channel width. 
    The $90^{\circ}$ reflection at the end of the U-base (indicated by the 
    vertical dashed line) has no influence on the symmetry of the distribution. 
  }
\end{figure}

These features can be understood with the sketches in Figure~\ref{fig:reflection-sketch}, which 
illustrate two cases of photocathode illumination for an electron traversing the channel along 
its central axis and assuming the Cherenkov angle of the chosen gas ($\Theta_{\rm Ch} = 3^{\circ}$).
Photons emitted in the horizontal (or vertical) plane illuminate the entire width of the channel 
at the photocathode as shown in Figure~\ref{fig:reflection-sketch}(a). Due to the larger effective 
channel cross section, photons emitted towards the corners illuminate only half the channel width 
at the photocathode as illustrated in Figure~\ref{fig:reflection-sketch}(b). 
This leads to a higher photon yield near the diagonals of the channel cross section which also 
explains the X-like structure observed in Figure~\ref{fig:LightYield-comp}. 
In addition, Figure~\ref{fig:reflection-sketch}(b) explaines the fact that the lower reflectivity 
of the inter-channel wall leads to a depletion on the same side of the channel.

\subsection{Yield asymmetries and beam position} \label{sub:sim-asym}
\begin{figure}[!h]
  \begin{picture}(16.0, 13.5)
    \put(-0.05, 8.40)  {\epsfig{file=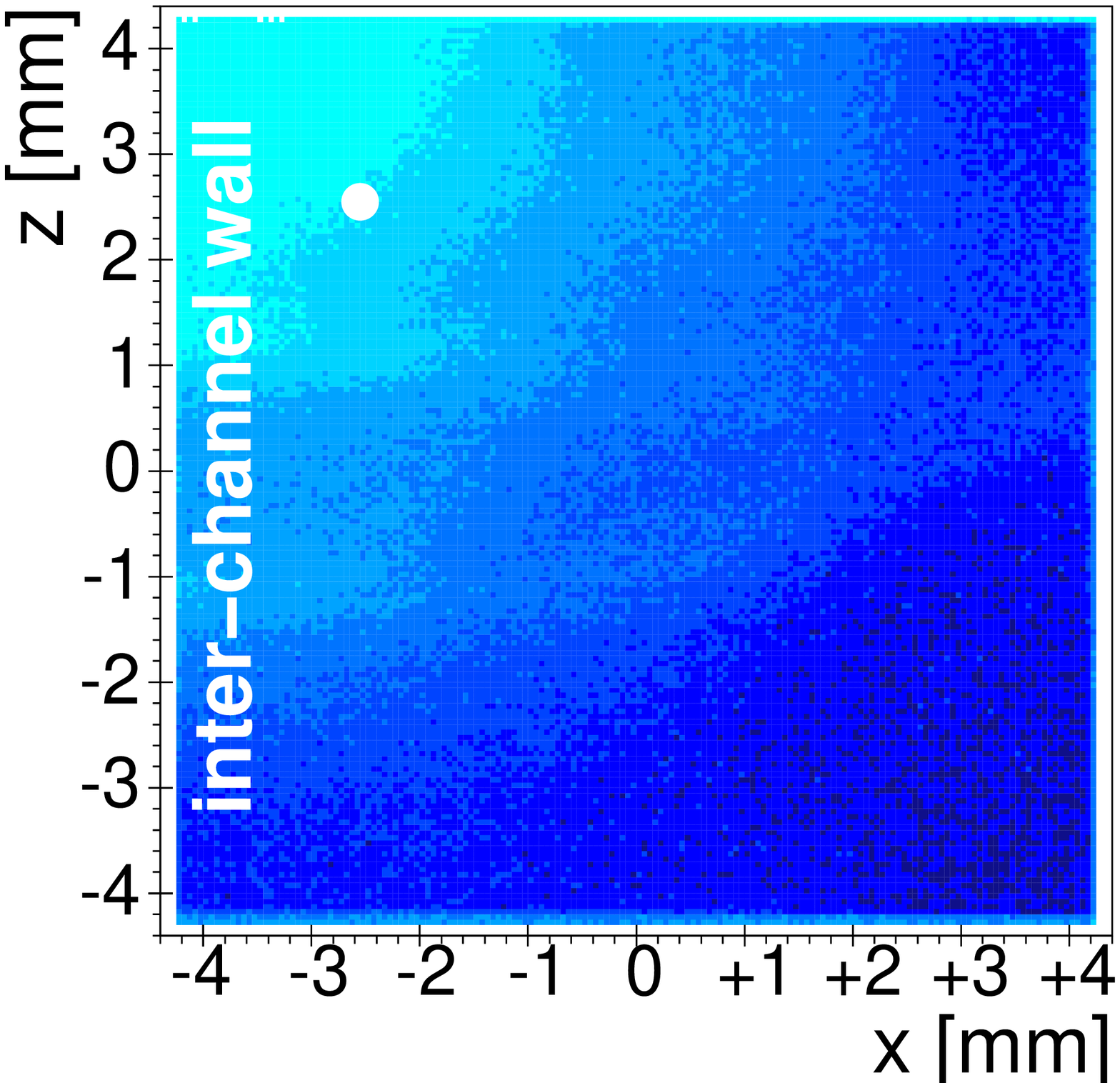,  clip= , height=0.26\linewidth}}
    \put( 4.15, 8.40)  {\epsfig{file=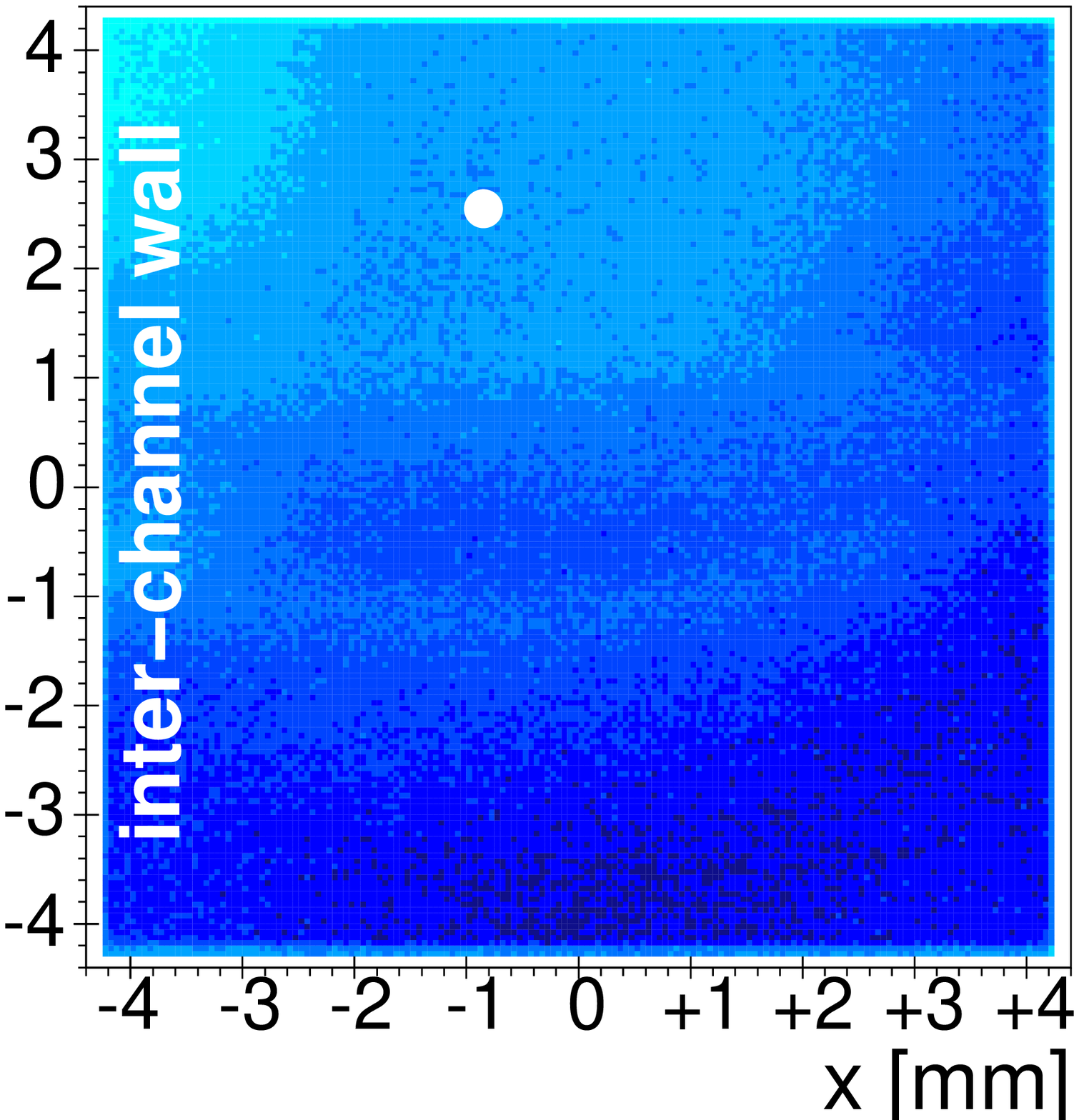,  clip= , height=0.26\linewidth}}
    \put( 8.03, 8.40)  {\epsfig{file=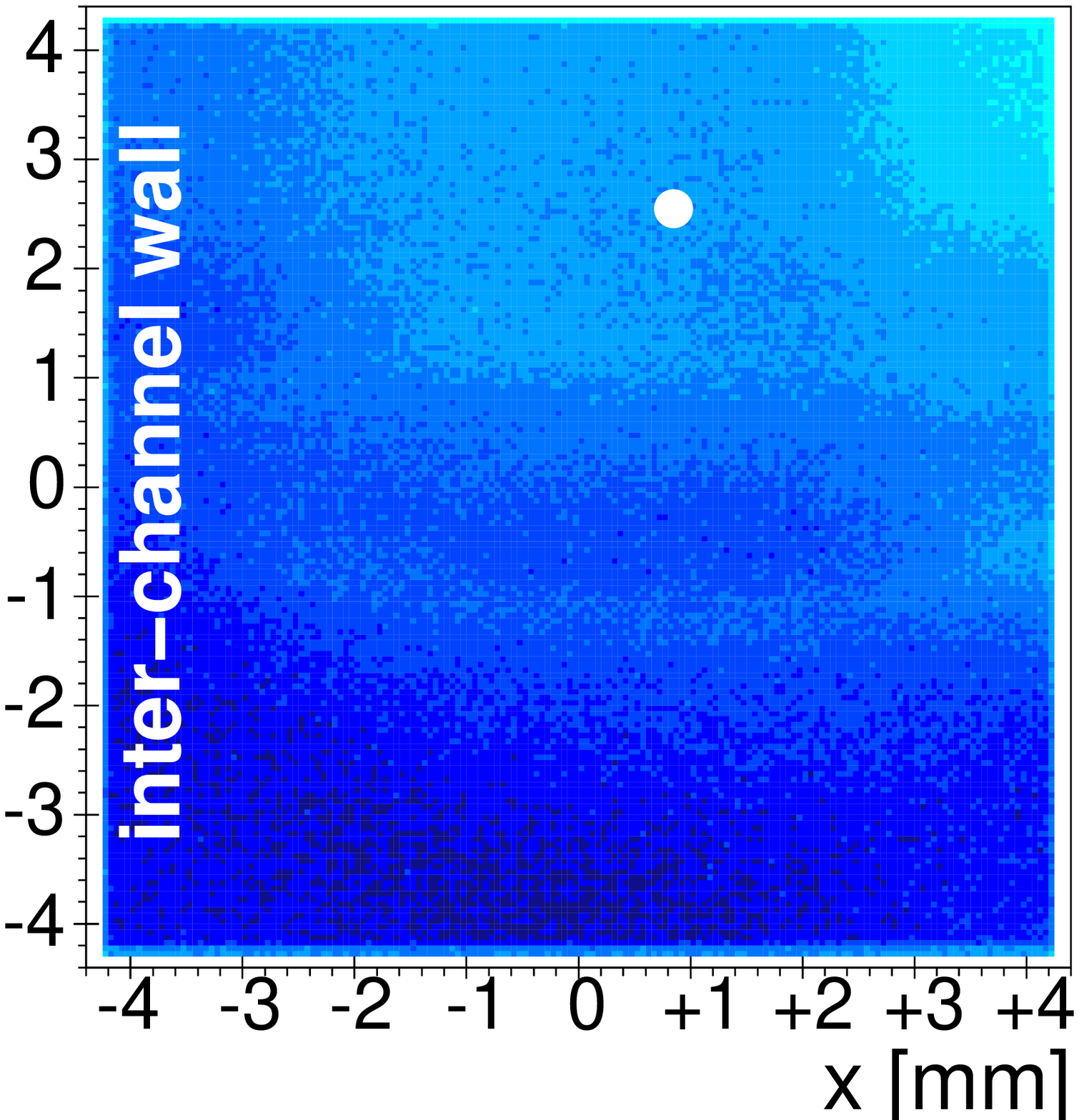,  clip= , height=0.26\linewidth}}
    \put(11.90, 8.40)  {\epsfig{file=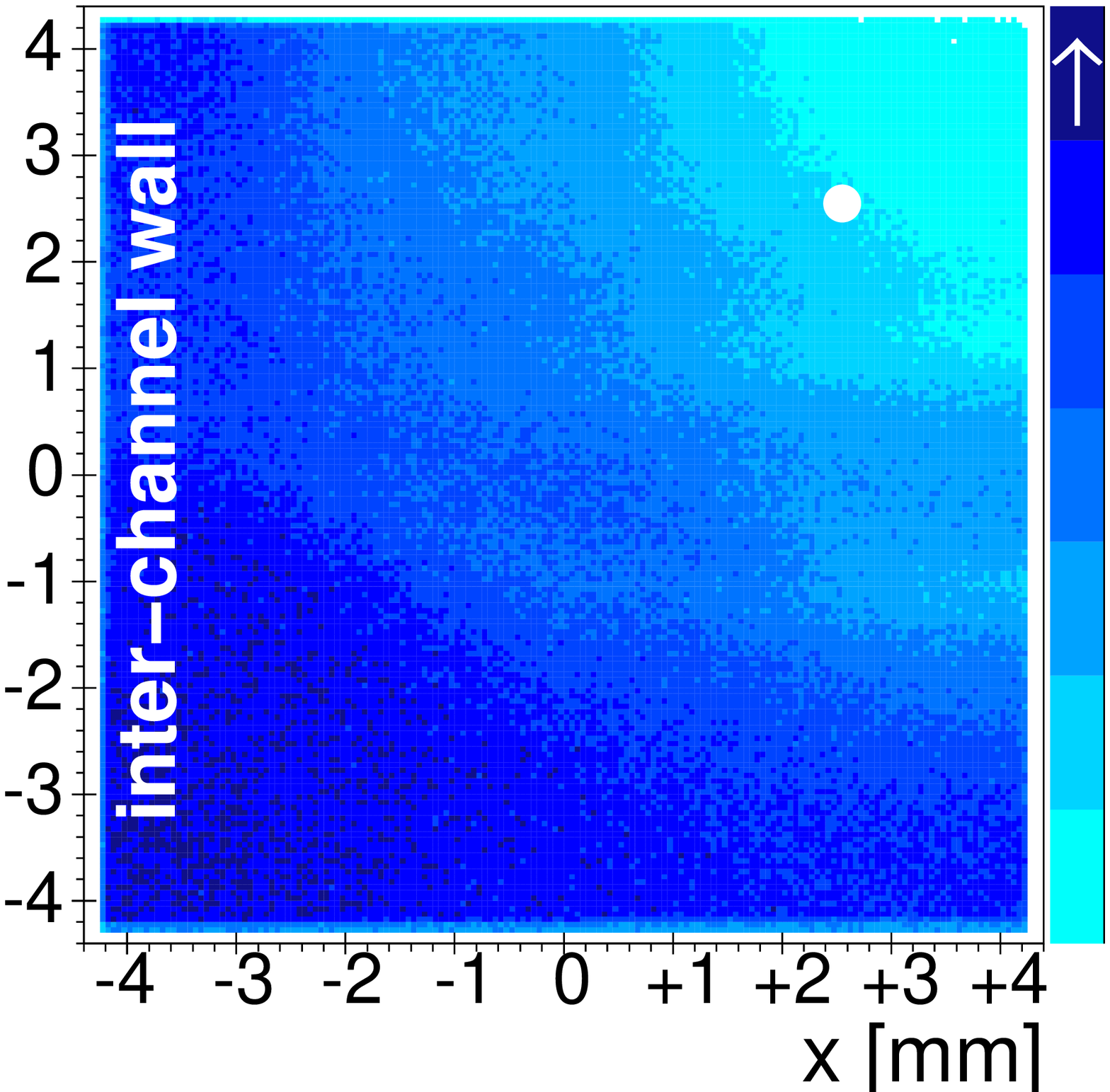,  clip= , height=0.26\linewidth}}
    \put(-0.05, 0.00)  {\epsfig{file=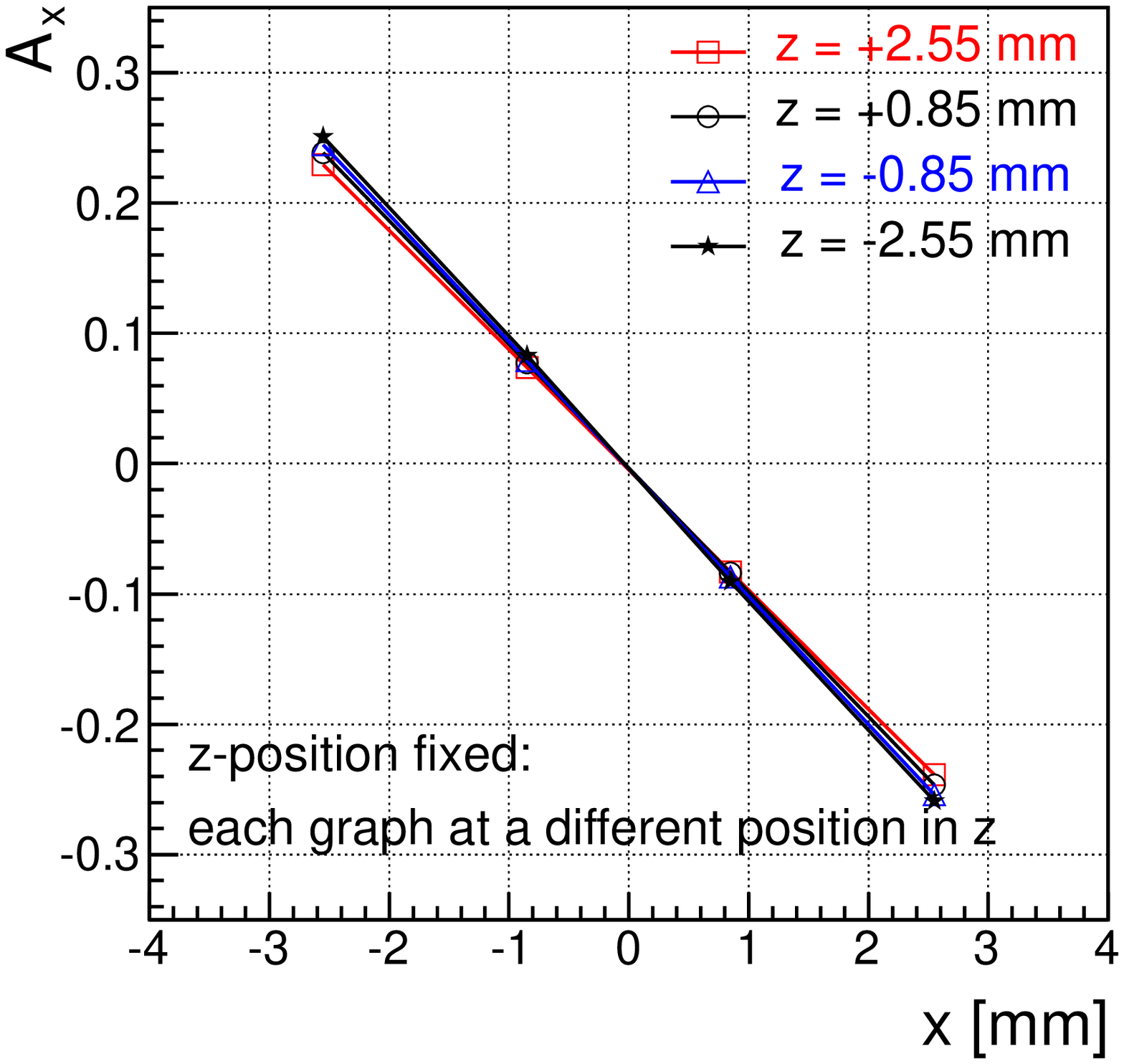, clip= ,  width=0.50\linewidth}}
    \put( 8.15, 0.00)  {\epsfig{file=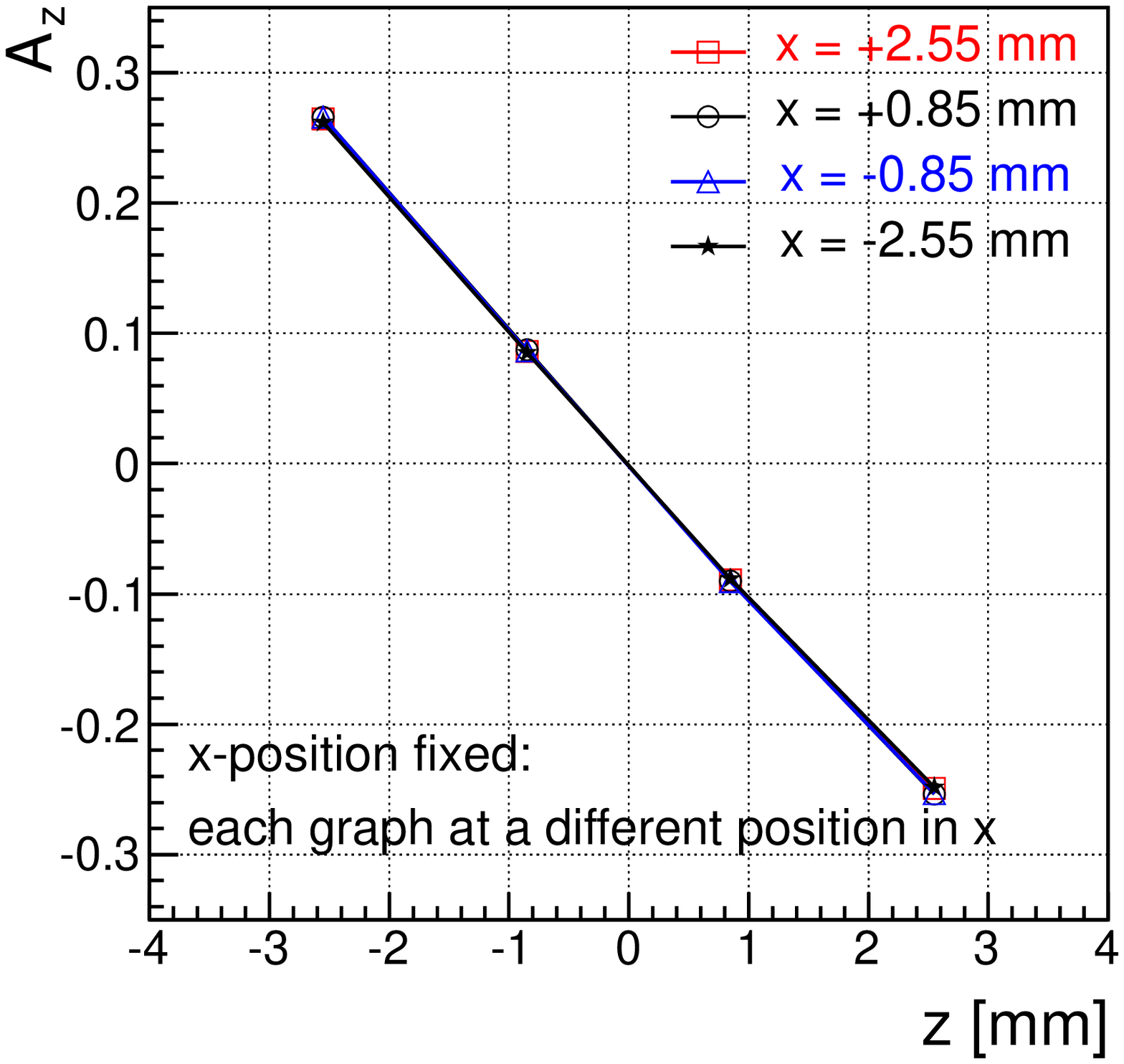, clip= ,  width=0.50\linewidth}}
    \put( 0.00, 8.30)  {(a)}
    \put( 0.00, 0.00)  {(b)}
    \put( 8.30, 0.00)  {(c)}
  \end{picture}
  \caption{\label{fig:SimGridScan-AsymXZ} \it 
    (a) Simulated light yield at the photocathode for a horizontal beam position scan at 
    $y\!=\!+2.55\:\milli\meter$ and equal reflectivities for all channel walls.
    The asymmetries have been calculated respectively from horizontal and 
    vertical scans using $10^5$~electrons for each beam position: 
    (b)~$A_x$ asymmetry for different $z$-positions and 
    (c)~$A_z$ asymmetry for different $x$-positions. \newline
      The beam $y$-position translates directly to the $z$-position in the readout plane
      (white dots).}
\end{figure}

The light distribution on the photocathode has been simulated for a grid scan of $4\!\times\!4$ 
beam positions with $10^5$ electrons per position, assuming equal reflectivities for all channel walls. Figure~\ref{fig:SimGridScan-AsymXZ}(a) depicts 
the light yield on the photocathode for electrons entering the Cherenkov section at a fixed 
$y$-position and four different $x$-positions.  
The $y$-position of the beam, indicated by the white dots in Figure~\ref{fig:SimGridScan-AsymXZ}(a), 
translates directly to the $z$-position in the readout plane of the photodetector. A clear correlation between the light pattern and the beam position can be observed in these simulations. If this correlation persists in real data, it could potentially be a very useful tool both in a testbeam experiment for tuning the simulation and studying the detector properties in detail, as well as in the final polarimeter for monitoring the alignment and the Compton edge position during data-taking without the need for expensive dedicated calibration beam time.

In order to pursue this idea further, the light pattern is quantified in terms of the 
asymmetries in Figures~\ref{fig:SimGridScan-AsymXZ}(b,c), which are calculated from the light  intensities for scans in the $x$- and $z$-directions, respectively. For each beam position the corresponding asymmetries are defined as

\begin{displaymath}
  \mbox{\large$A_x$} = \mbox{\LARGE $\frac{ I_x^+ - I_x^- } { I_x^+ + I_x^- }$} \hspace*{10mm} \mbox{ and } \hspace*{10mm} 
  \mbox{\large$A_z$} = \mbox{\LARGE $\frac{ I_z^+ - I_z^- } { I_z^+ + I_z^- }\,$,} 
\end{displaymath}

where $I_x^+$ ($I_z^+$) corresponds to the intensity in the right (upper) half of a channel 
and   $I_x^-$ ($I_z^-$) to the intensity in the left (lower) half, respectively.

These asymmetries have an approximately linear dependence on the beam position. 

Figure~\ref{fig:SimGridScan-AsymXZ}(b) shows a slight variation of the slope of $A_x$ depending on the $y$ position of the beam, while in Figure~\ref{fig:SimGridScan-AsymXZ}(c) the same $A_z$ is observed for all $x$ positions. This behavior is expected since the active length of the channels increases with $y$ due to the $45^{\circ}$ angle of the mirrors.

\subsection{Reflectivity measurements} \label{sub:sim-reflect}
When choosing the detector materials, different qualities of aluminium have been considered, 
primarily with regard to their reflectivity, but also concerning smoothness and mechanical stability.

\begin{figure}[!h]
  \begin{picture}(16.0, 7.5)
    \put( 0.00, 0.20)  {\epsfig{file=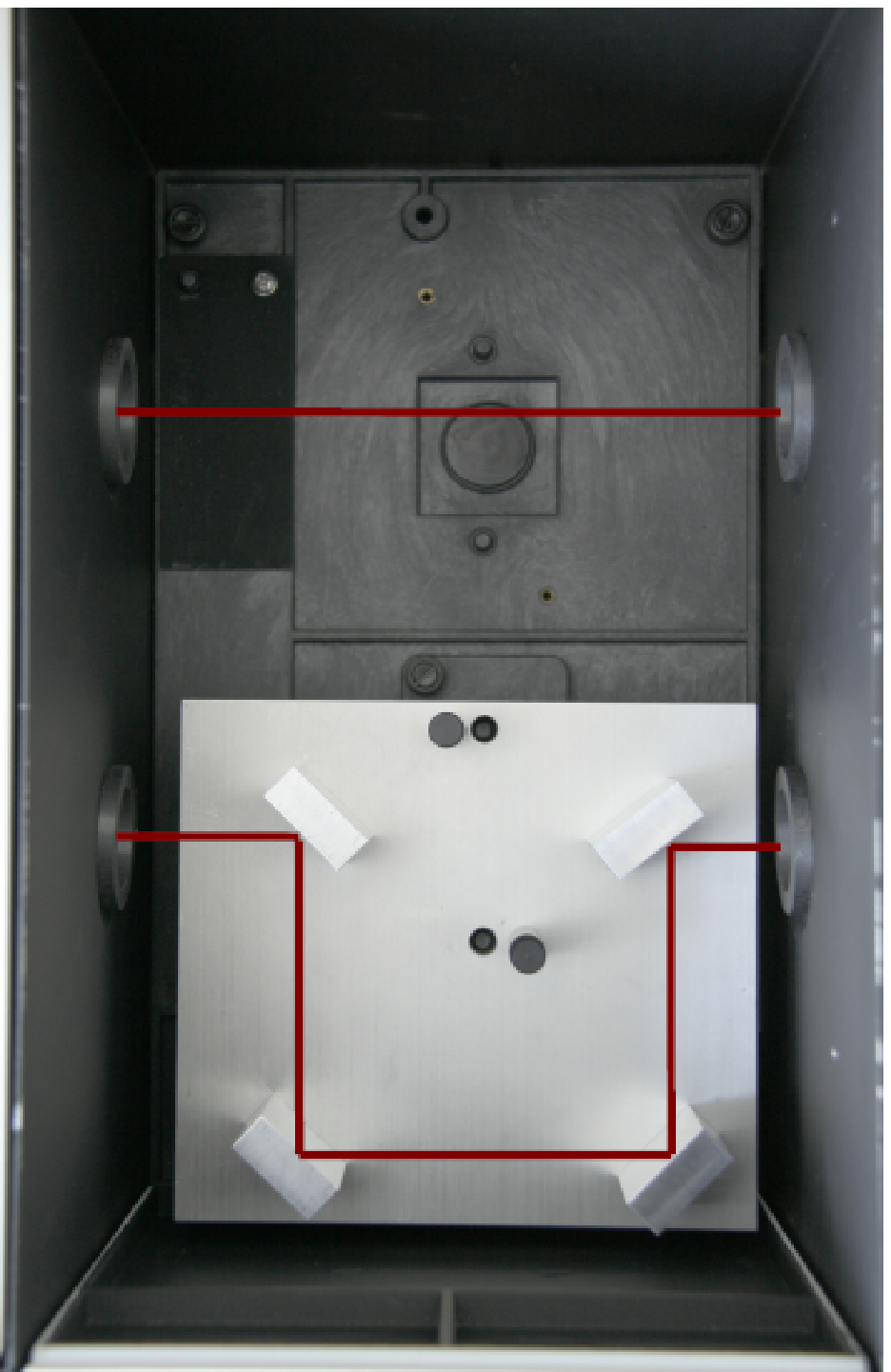,          clip= , width=0.27\linewidth}}
    \put( 4.65,-0.25)  {\epsfig{file=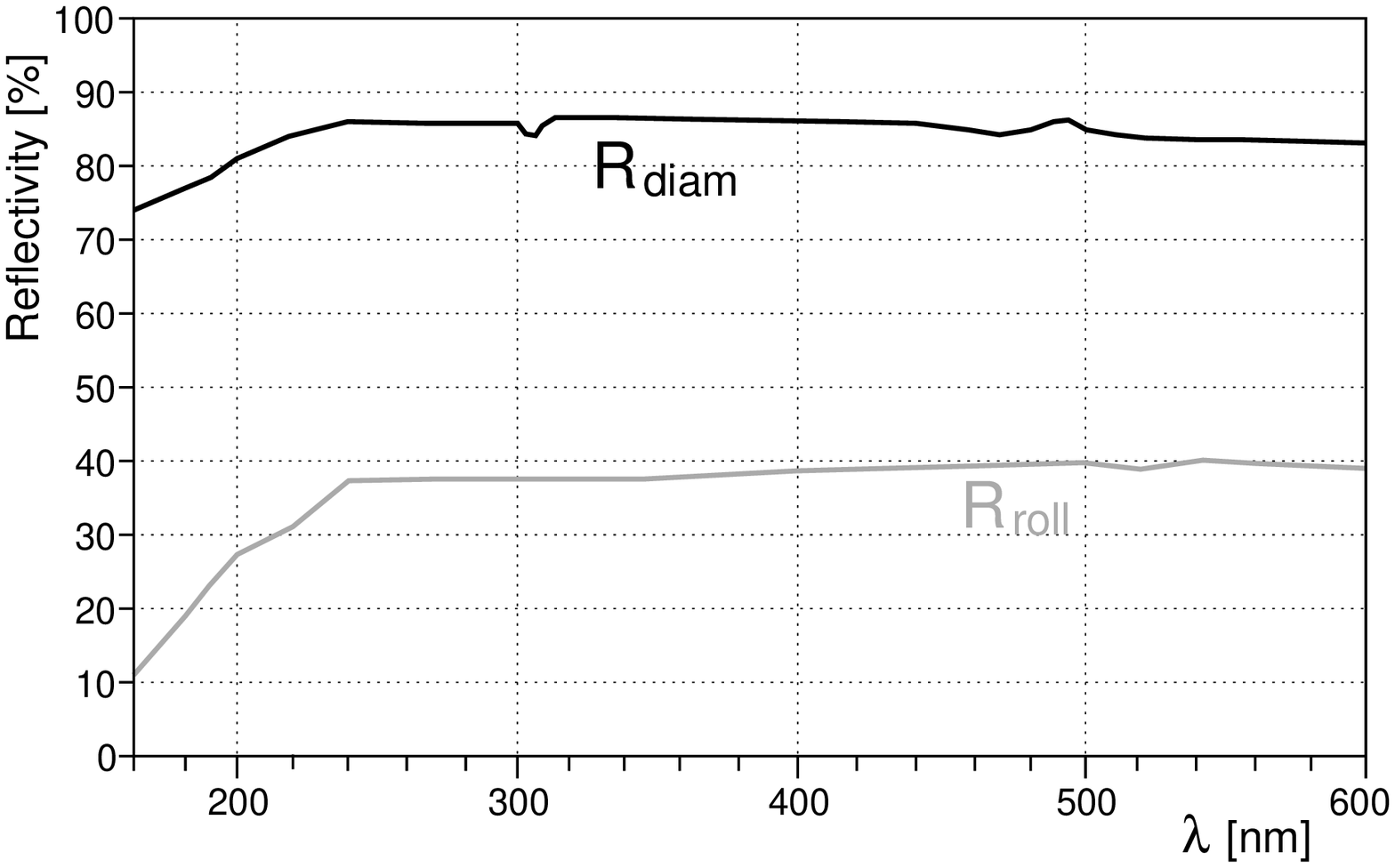, clip= , width=0.72\linewidth}}
    \put( 0.25, 6.30)  {\bf\textcolor{white}{(a)}}
    \put(15.00, 6.20)  {(b)}
  \end{picture}
  \caption{\label{fig:PerkinElmer-Refl} \it 
    (a) Photograph of the interior of the modified \perkinelmer{} transmission spectrometer 
    with the indicated paths of the reference beam (top) and the measurement beam (bottom).
    (b) Measured reflectivities of diamond-milled aluminium (upper line) 
    and of rolled aluminium (lower line).
  }
\end{figure}

Reflectivity measurements of small aluminium probes (blocks, sheets, and sub~mm-foils) of different 
quality have been performed with a modified transmission spectrometer~\cite{bib:PerkinElmer}.
The path of the measurement beam inside the spectrometer has been changed such that it reflects off 
four small blocks instead of passing through the probe material, as shown in Figure~\ref{fig:PerkinElmer-Refl}(a). 
One photomultiplier detects the previously splitted reference and measurement beams and thus 
provides a measure of how much light is reflected by the four blocks' surfaces with respect to
the reference beam. 

Figure~\ref{fig:PerkinElmer-Refl}(b) shows measurements of the reflectivity as a function of the wavelength between $\lambda = 160$~nm and $600$~nm for diamond-milled aluminium blocks and for a sample of rolled aluminium foil purchased from~\goodfellow\footnote{\goodfellow{} GmbH, Germany; Aluminium foil: AL000601 (thickness: 0.15~mm, purity: 99.0\%, hardness: hard)}. In both cases the reflectivity rises at low wavelengths up to $\lambda \approx 250$~nm and is approximately constant at larger wavelengths. For the diamond-milled blocks, the reflectivity reaches $R_{\rm diam} \approx 85\%$ in the plateau region, which is in good agreement with~\cite{bib:AlRef}. This material is employed for the mirrors at the ends of the U-base, where the reflection occurs approximately under $45^{\circ}$ as in the spectrometer. Therefore these measured values have been implemented in the simulation.

\begin{figure}[!h]
  \begin{picture}(7.6, 7.6)
    \put(-0.25, 0.00)  {\epsfig{file=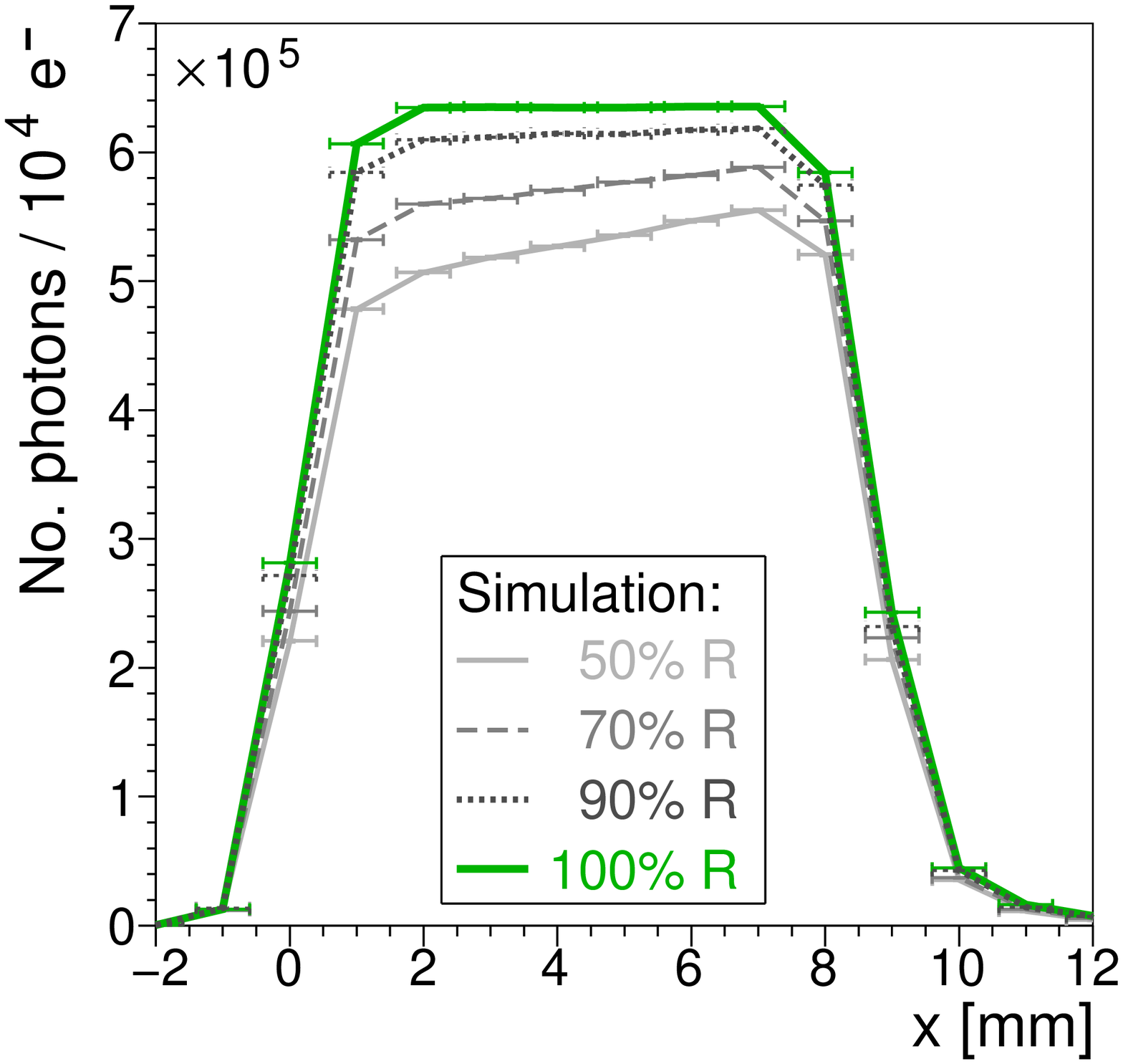,     clip= , width=0.51\linewidth}}
    \put( 8.00, 0.00)  {\epsfig{file=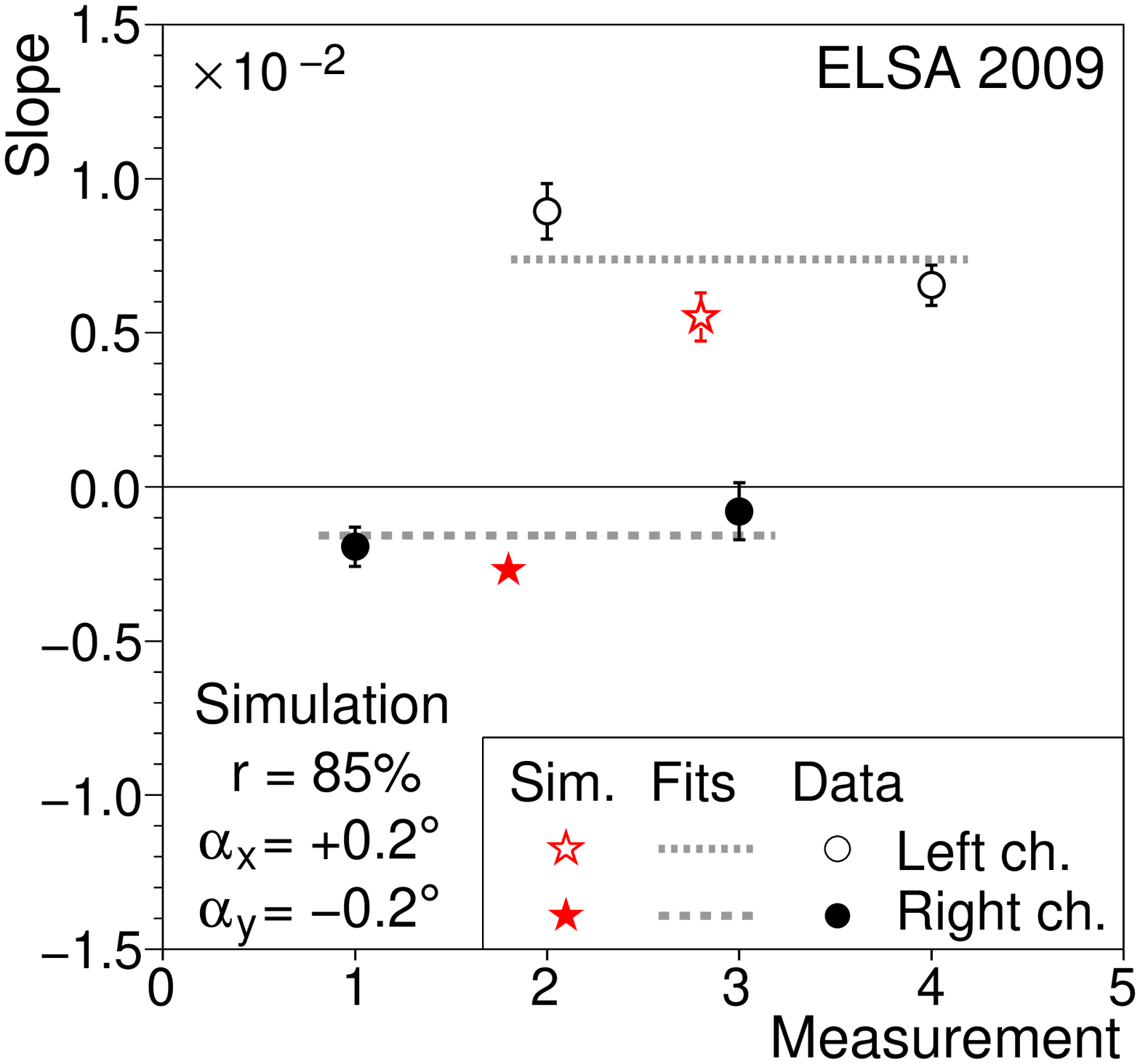, clip= , width=0.51\linewidth}}
    \put( 0.00, 0.00)  {(a)}
    \put( 8.25, 0.00)  {(b)}
  \end{picture}
  \caption{\label{fig:xscan-sim-varref} \it
    (a) 
    Simulations with different percentages of the inter-channel wall 
    reflectivity w.r.t.\ to the other walls' reflectivities lead to 
    different plateau shapes. 
    (b) A comparison with data allows to determine the inter-channel wall 
     reflectivity to be $\,R_{roll}= (85\% \pm 2\%) \cdot R_{diam}$ under glancing angle.
  }
\end{figure}

In case of the rolled foil, the measured plateau value of $R_{\rm roll} \approx 40\%$ is low, but still in the range of reflectivities observed for not perfectly smooth surfaces~\cite{bib:AlRef}. However it is not directly applicable to the situation in the prototype for several reaons: To start with, the reflections on the channel walls occur predominantly under shallow angles ($\Theta_{\rm Ch} = 3^{\circ}$). 
Furthermore, any light not following the path expected for an ideal reflector, be it due to a finite size of the specular reflection highlight or due to a diffuse reflection component, will lead to an immediate intensity loss in the spectrometer, while it still has a high probability to reach the photodetector at end of the channel in the prototype. Therefore the absolute scale of the spectrometer measurements is considered as a lower limit for the effective reflectivity of the inter-channel wall. The wavelength dependency observed in the spectrometer measurements is taken into account in the simulation, but the overall ratio  $r = R_{\rm roll}^{\rm eff}/R_{\rm diam}$ is a parameter which is varied between 50\% and 100\% as extreme cases, and which ultimately needs to be determined from prototype data. In particular, it can be determined from the detector response as function of the beam position if enough position scans with sufficient beam quality are available. This is illustrated in Figure~\ref{fig:xscan-sim-varref}(a), where simulated beam position scans for different values for $r$ exhibit different slopes of the plateau region towards the inter-channel wall. Figure~\ref{fig:xscan-sim-varref}(b) shows measurements of this slope for both detector channels. The different absolute values of the slopes observed for the left and right channel are due to residual misalignment (c.f. Section~\ref{sub:tb-asym}). Comparisons with a simulation including the misalignment suggest a realistic value of $r$ of approximately 85\%, which was used in further simulations.

\section{Construction of the Prototype}  \label{sec:construction}
The channel dimensions of the prototype were chosen to match the design criteria discussed 
in section~\ref{sub:sim-concept}. The length of the U-base relevant for the emission of 
Cherenkov radiation from traversing electrons is $150\:\milli\meter$ and the height of the 
two U-legs is $100\:\milli\meter$. 
A quadratic cross section of $8.5\times8.5\:\milli\meter^2$ has been chosen to match the 
cathode geometry of two square multi-anode photomultipliers. Section~\ref{sub:const-photodet} 
gives further details of the employed photodetectors and their characteristics. 

\begin{figure}[!h]
  \begin{picture}(16.0, 11.5)
    \put(-0.80, 0.20)  {\epsfig{file=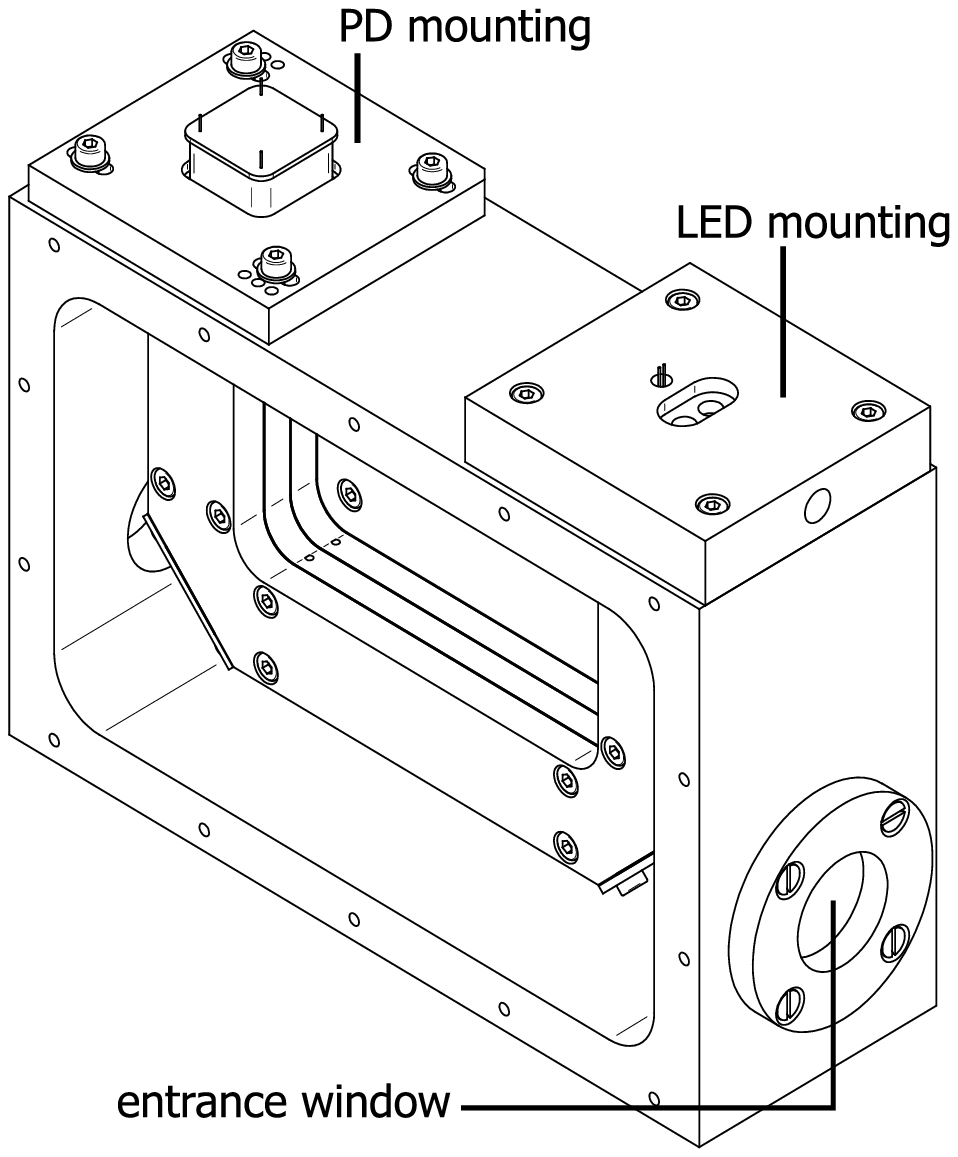, width=0.58\linewidth}}
    \put( 8.80, 1.20)  {\epsfig{file=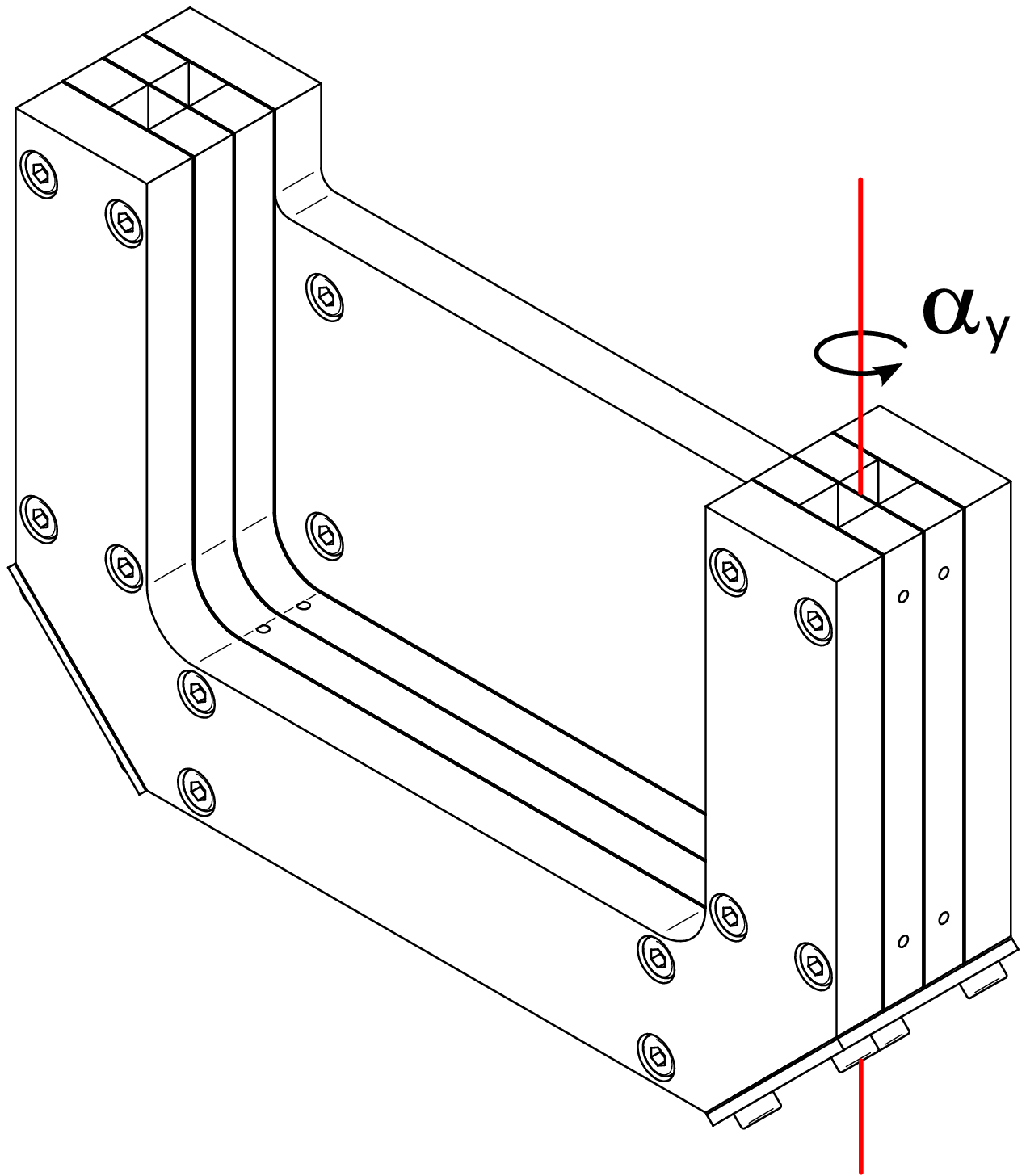,  width=0.45\linewidth}}
    \put( 0.00, 0.20)  {(a)}
    \put( 9.20, 0.20)  {(b)}
  \end{picture}
  \caption{\label{fig:TechnicalDrawings} \it 
    Parts of a technical drawing for the assembly of the prototype: 
    (a) the box base body, already including the channel structure 
    (b) of two parallel U-shaped channels.
  }
\end{figure}

The size of the outer box is $230 \times 90 \times 150\:\milli\meter^3$~(L$\times$W$\times$H) 
allowing for easy accomodation of the channel structure. Parts of the technical drawing, e.g.\ 
the channel structure and its placement inside the box, are shown in Figure~\ref{fig:TechnicalDrawings}. 

Perfluorobutane was chosen as Cherenkov gas due to its high Cherenkov threshold of about 
$10\:\mega\electronvolt$ for electrons. In addition it is neither flammable, nor explosive, 
contrary to propane or isobutane. The $10\:\milli\meter$ thick aluminium lid of the box holds 
an electronic pressure gauge suited for remote read-out. The entrance and exit windows for 
the electron beam consist of $0.5\:\milli\meter$ thin aluminium sheets. 

All mountings for LEDs, photodetectors and windows have been designed to be gas- and 
light-tight, as well as easily exchangeable.

\subsection{Photodetectors and their mountings} \label{sub:const-photodet}
\begin{figure}[!h]
  \begin{picture}(16.0, 5.2)
    \put( 0.00, 0.50)  {\epsfig{file=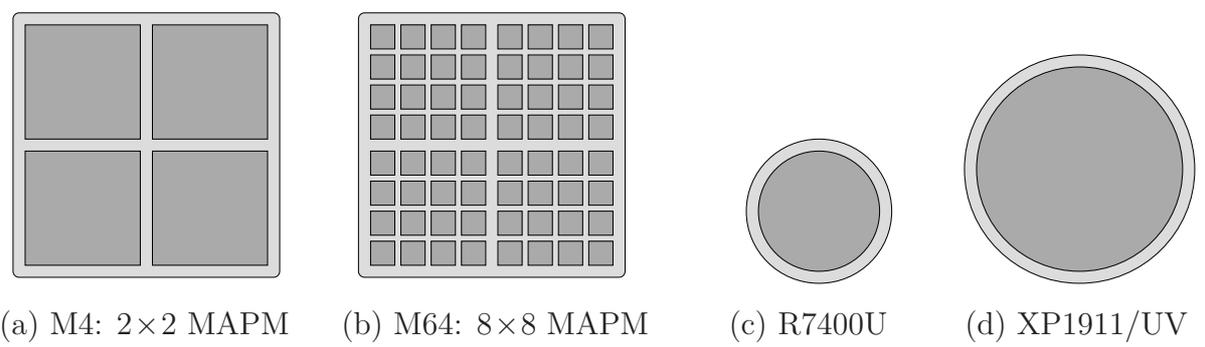, clip= , width=1.00\linewidth}}
    \put( 0.00, 0.00)  {(a) M4:  $2\!\times\!2$~MAPM }
    \put( 4.50, 0.00)  {(b) M64: $8\!\times\!8$~MAPM }
    \put( 9.60, 0.00)  {(c) R7400U }
    \put(12.70, 0.00)  {(d) XP1911/UV }
  \end{picture}
  \caption{\label{fig:PM-AnodeSchemes} \it 
    Anode schemes of the different multi- and single-anode 
    photomultiplier types, in correct relative scaling.
  }
\end{figure}

The hind U-leg can be equipped with four different types of photodetectors, which 
are listed in Table~\ref{tab:PM-types} along with some of their characteristics. 
They differ in geometry (square versus round) and the number of anode pads as 
illustrated in Figure~\ref{fig:PM-AnodeSchemes}. 
Their gains are in the order of $10^6$ with wavelength thresholds between $160\:\nano\meter$ 
and $300\:\nano\meter$ and their response times range from $6.5\:\nano\second$ to $28\:\nano\second$. 
In case of the square multi-anode photodetectors (MAPMs), one quadrant of their cathodes 
exactly matches one detector channel. Thus, both detector channels can be read out 
simultaneously by the same photodetector. 

\begin{figure}[!h]
  \begin{picture}(16.0, 4.5)
    \put( 0.00, 0.40)  {\epsfig{file=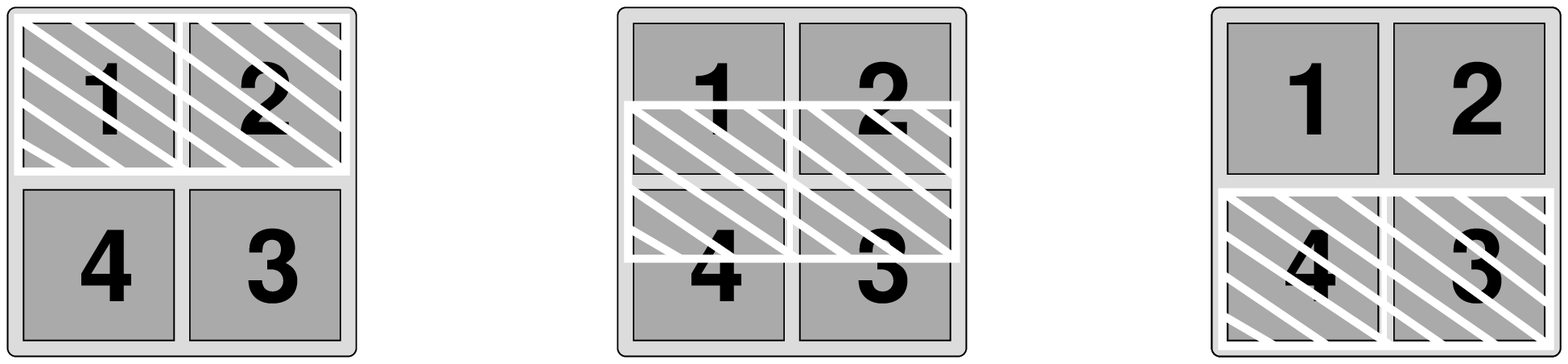, clip= , width=1.00\linewidth}}
    \put( 0.20, 0.00)  {(a) upper position}
    \put( 6.30, 0.00)  {(b) middle position}
    \put(12.20, 0.00)  {(c) lower position}
  \end{picture}
  \caption{\label{fig:MountingPos} \it 
    Different positions of the MAPMs (grey) on the detector channels (hatched).
  }
\end{figure}

While the round single-anode photodetectors (SAPMs) are inserted into their respective 
mountings using appropriate O-ring seals, the MAPMs need to be glued into 
their mountings. Epoxy resin mixed with black paint was used as glue to ensure gas- and also  
light-tightness. The mountings themselves were manufactured from poly-oxy-methylene (POM) 
for electrical insulation. They provide for three different photodetector positions 
relative to the detector channels as depicted in Figure~\ref{fig:MountingPos}.
In addition, both MAPM mountings can be rotated by $180^{\circ}$ for systematic studies.

\subsection{LED calibration system}\label{sub:const-LEDcalib}
The front U-leg of the detector serves for calibration purposes and is equipped with 
one LED per channel. The LEDs have a peak wavelength of $470\:\nano\meter$~(HLMP-CB30-NRG, 
Agilent Technologies~\cite{bib:LEDs-AgilentTech}) and are glued into their mounting 
structure using epoxy resin. 

\begin{figure}[!h]
  \begin{picture}(16.00, 5.00)
    \put( 5.60, 0.00)  {\epsfig{file=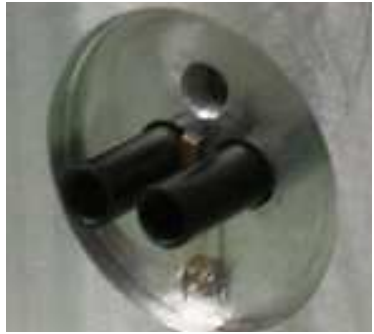, bb=0 645 110 745, clip= , width=0.30\linewidth}}
  \end{picture}
  \caption{\label{fig:LED-tubes} \it  Calibration LEDs covered by POM tubes.}
\end{figure}

As shown in Figure~\ref{fig:LED-tubes}, two slender $18\:\milli\meter$ long POM tubes 
encase the LEDs to ensure that the light from one LED does not enter the neighbouring 
channel through a small slit in the inter-channel wall necessary for gas circulation. 
A temperature sensor is placed in between the two POM tubes to allow for temperature 
monitoring.

The LEDs and the temperature sensor are fixed in a mounting which has been designed to 
be easily exchangeable, because the prototype will serve in the future as a test bed for
a suitable calibration system which is currently under development.

\subsection{Additional components} \label{sub:const-components}
Two small, light-tight boxes protect the MAPMs and their electrical bases. 
A rotation mechanism on a plastic base plate allows to adjust the detector's 
horizontal tilt about the $y$-axis in reproducible steps 
of $0.125^{\circ}$ between $\alpha_y = \pm 3.0^{\circ}$. 
The fixed rotational axis lies in the center of the front U-leg as illustrated 
in Figure~\ref{fig:TechnicalDrawings}(b).

\section{Beam Tests at the ELSA Accelerator}  \label{sec:testbeam}
\begin{figure}[!h]
  \begin{picture}(16.0, 7.5)
    \put( 0.00, 0.00)  {\epsfig{file=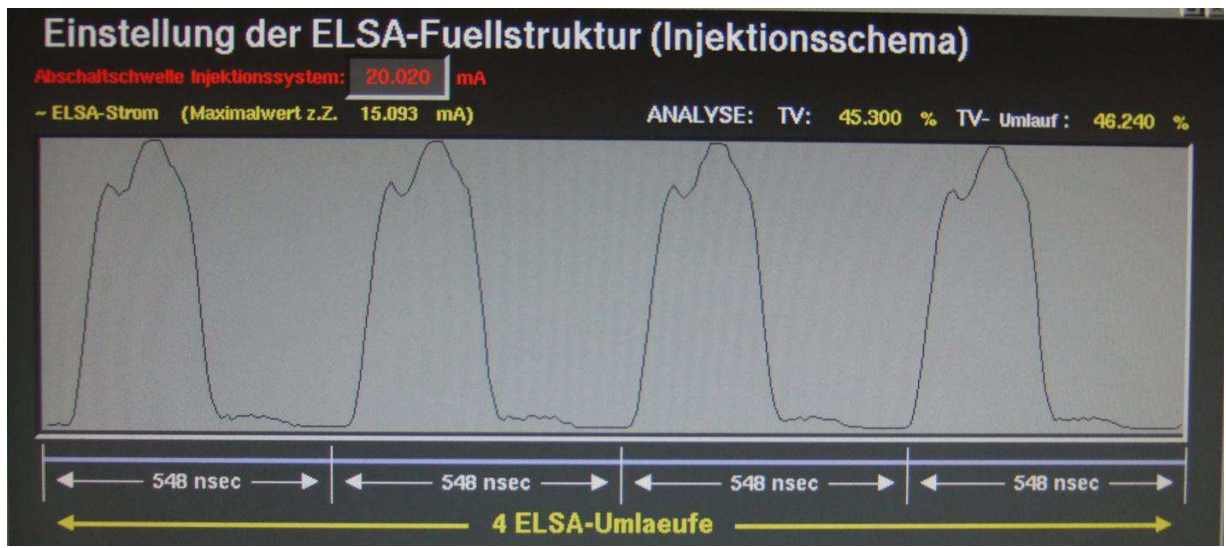, clip= , width=1.00\linewidth}}
  \end{picture}
  \caption{\label{fig:ELSA-FillStruct} \it 
    Example of a typical ELSA fill structure for four 
    revolutions of  $548\:\nano\second$.
    About half of the available buckets are filled.
  }
\end{figure}

Beam tests with the prototype detector were performed in an external beam line at ELSA. 
The {\sc EL}ektronen-{\sc S}tretcher-{\sc A}nlage (ELSA) is an electron accelerator consisting 
of three stages: injector LINACs, a booster synchrotron and the stretcher ring\cite{Hillert:2006yb}. 
A beam of polarised or unpolarised electrons of variable energy up to $3.5\:\giga\electronvolt$ can 
be stored and used for various experiments in different beam line areas around the storage ring. 
The stretcher ring has a circumference of $164.4\:\meter$ corresponding to a 
time of $548\:\nano\second$ for one revolution.
The ELSA beam is structured by the RF frequency of $500\:\mega\hertz$.  
Of $274$ buckets in total, a variable fraction can be filled. 
As an example, Figure~\ref{fig:ELSA-FillStruct} shows the fill structure for 
four revolutions of $548\:\nano\second$ for a partially filled ELSA accelerator.

\subsection{Setup and pedestal stability} \label{sub:tb-elsa}
During the testbeam period, ELSA was operated in {\it booster mode} with the electrons 
being injected at an energy of $1.2\:\giga\electronvolt$ and subsequently accelerated 
to $2.0\:\giga\electronvolt$. 
The machine cycle is $5.1\:\second$ with an extraction time of $4.0\:\second$ 
  and the beam can be focussed to a spot size of about $1\:\milli\meter$.
The extraction current is adjustable from approximately $10\:\pico\ampere$ to $200\:\pico\ampere$ 
leading to respectively $35$ to $700$~electrons traversing the detector per ELSA  revolution. 
In comparison, up to $250$~electrons per bunch crossing are expected in the most populated 
channel of a polarimeter Cherenkov detector at the ILC.

\begin{figure}[!h]
  \begin{picture}(16.0, 6.0)
    \put( 0.00, 0.00)  {\epsfig{file=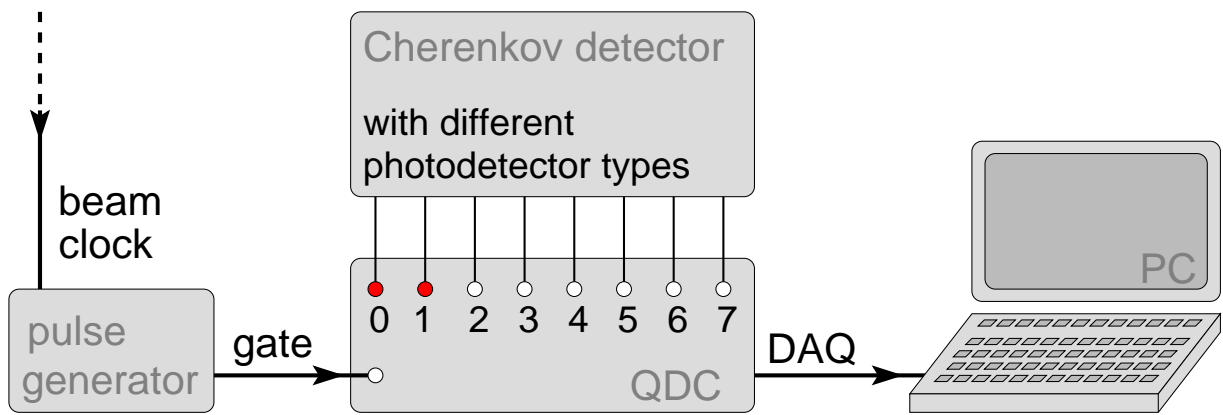, clip= , width=1.00\linewidth}}
  \end{picture}
  \caption{\label{fig:tb-readout} \it 
    Block diagram of the readout chain as realised during the testbeam period.
  }
\end{figure}

The beam clock signal was used to provide the gate for the QDC (charge sensitive 
analog-to-digital converter), as illustrated in the block diagram of the readout chain 
in Figure~\ref{fig:tb-readout}. The gate width was adjusted between $100\:\nano\second$ 
and $480\:\nano\second$ to integrate over the filled part of one ELSA revolution. 

\begin{figure}[!h]
  \begin{picture}(16.0, 13.5)
    \put( 3.40, 0.00)  {\epsfig{file=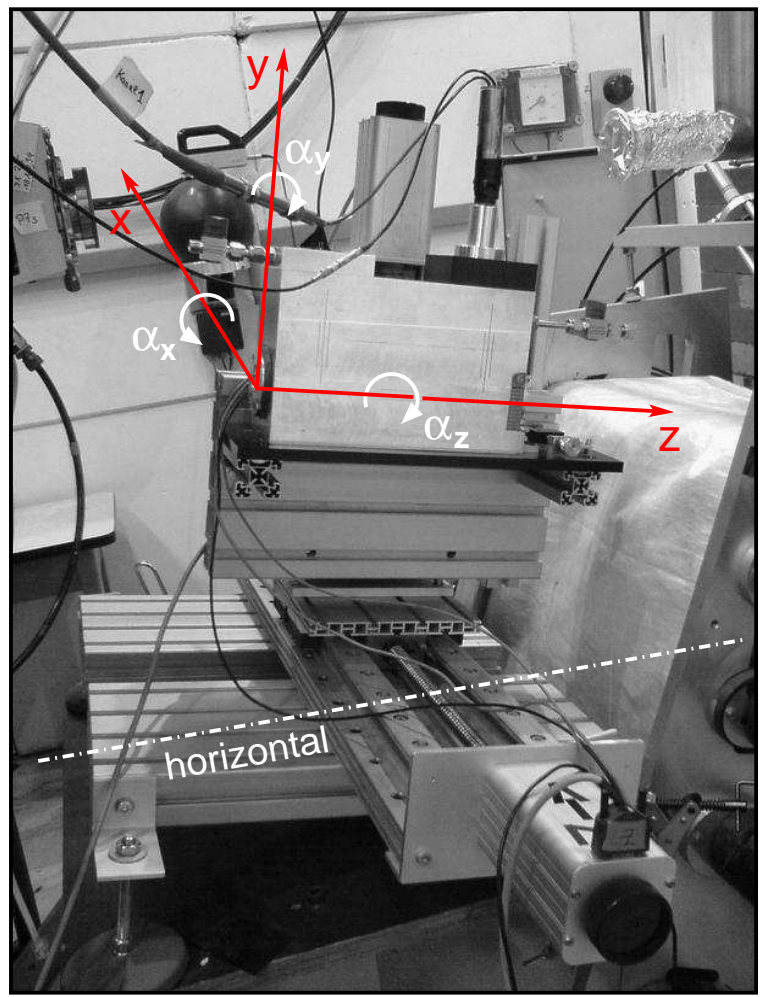, clip= , width=0.62\linewidth}}
  \end{picture}
  \caption{\label{fig:tb-setup} \it 
    The prototype Cherenkov detector on its base plate 
    (black, with the rotational mechanism visible on the right) 
    is mounted on a stage moveable along the $x$- and $y$-axis.
  }
\end{figure}

The detector was filled with the Cherenkov gas \CfFt{} at a slight overpressure of about 
$140\:\milli\mathrm{bar}$. This overpressure remained stable although frequent changes to 
the setup prevented a monitoring of the gas pressure for continuous time periods longer 
than two weeks. 
Figure~\ref{fig:tb-setup} shows the Cherenkov detector set up in one of the ELSA external 
beam lines, directly behind a dipole magnet bending the electrons by $\approx 7.5^{\circ}$ 
towards a downstream beam dump. The detector was mounted on its base plate (black) and 
additionally affixed to a stage moveable along the $x$- and $y$-axis. 
The two angles, $\alpha_x$ and $\alpha_z$, had to be adjusted using a water-level. 

\begin{figure}[!h]
  \begin{picture}(16.0, 8.0)
    \put(-0.05, 0.00)  {\epsfig{file=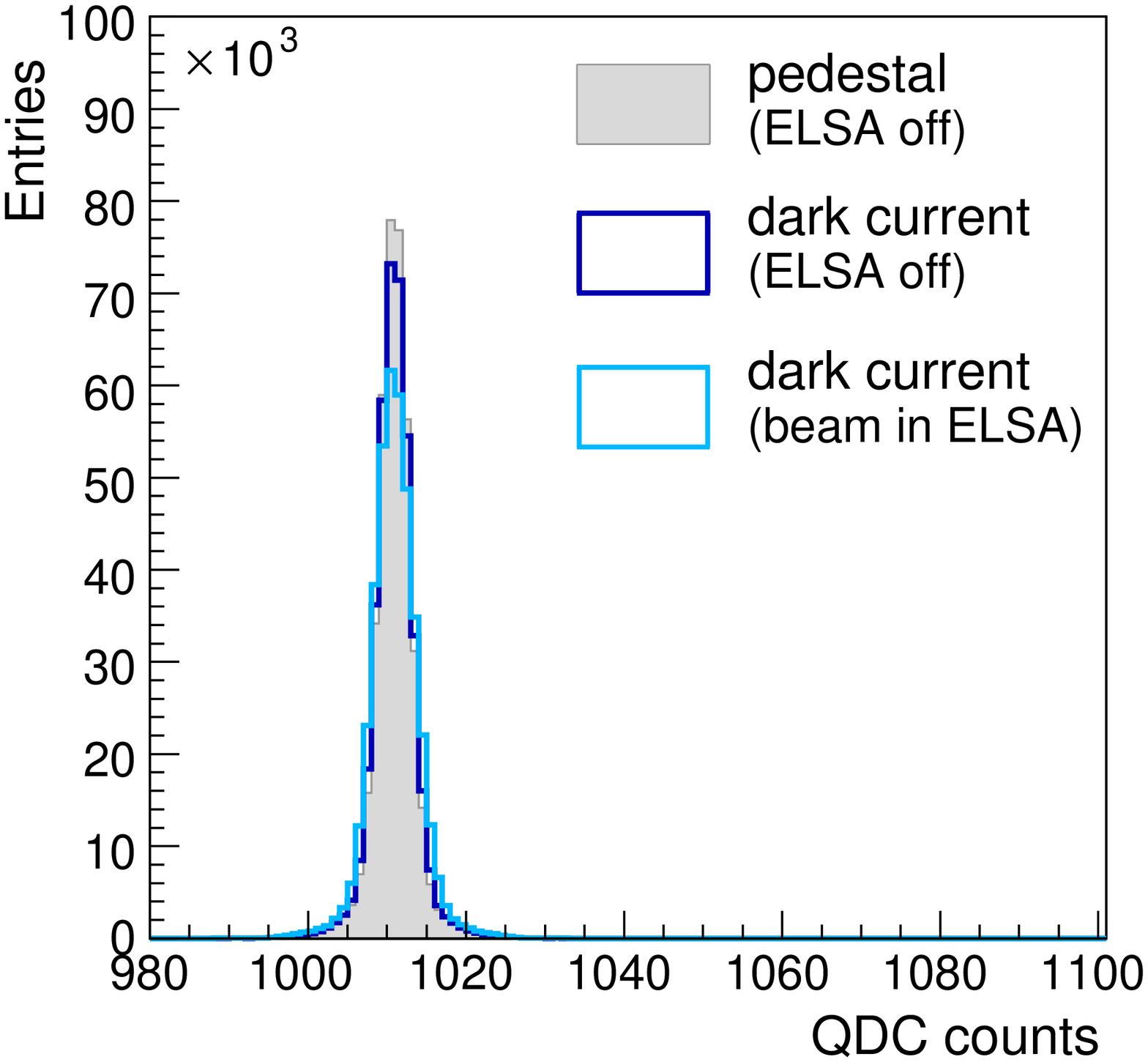, clip= , width=0.50\linewidth}}
    \put( 8.10, 0.00)  {\epsfig{file=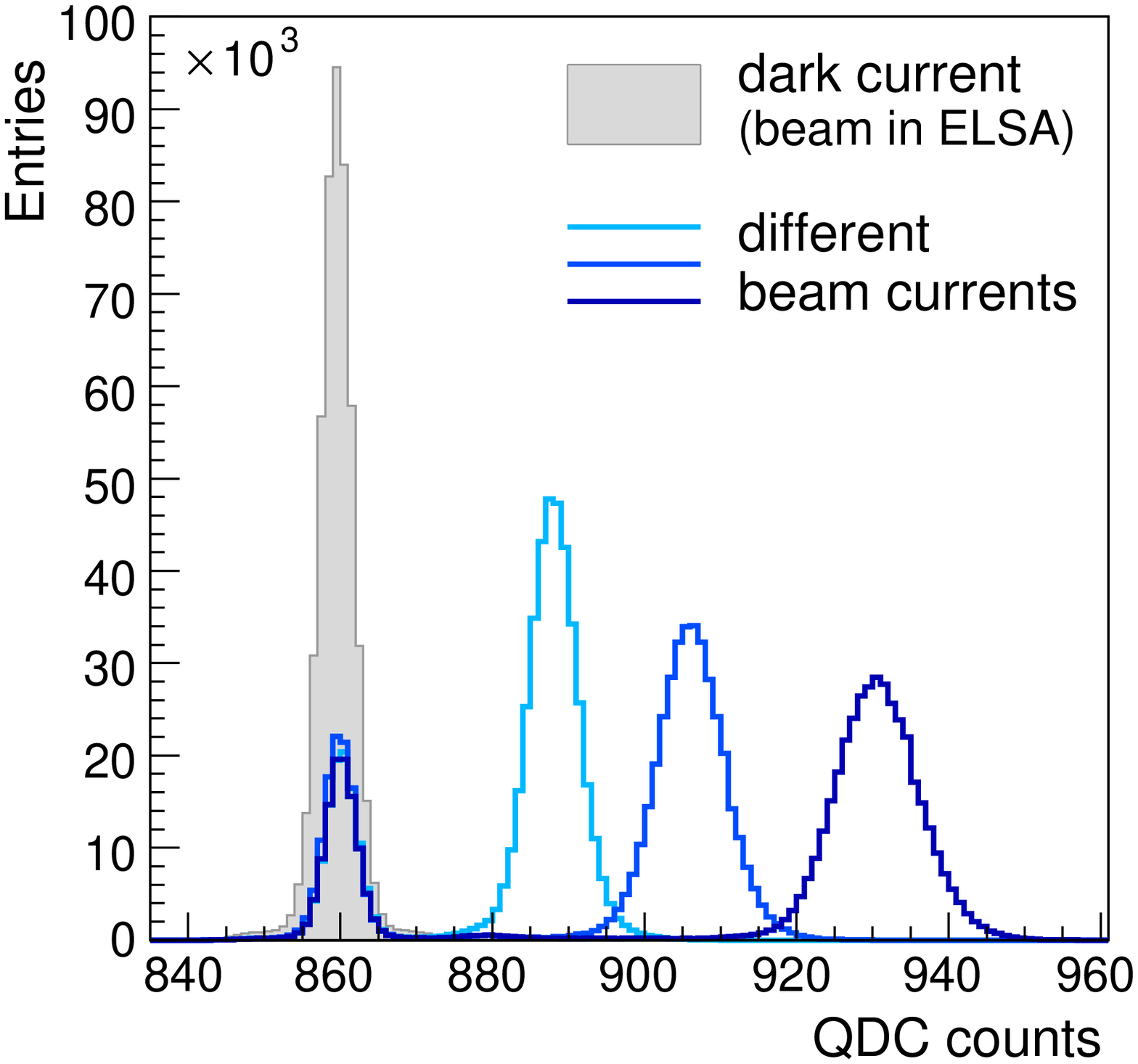, clip= , width=0.50\linewidth}}
    \put( 0.00, 0.00)  {(a)}
    \put( 8.35, 0.00)  {(b)}
  \end{picture}
  \caption{\label{fig:M4-Signal-DiffBC} \it 
    Data recorded with the $2\!\times\!2$ MAPM (R7600U-03-M4): \newline
    (a) QDC response without bias voltage applied to the photodetector (pedestal) and 
    with $400\:\volt$ applied (dark current), both without and with beam circulating in ELSA. \newline
    (b) Cherenkov signals increase with increasing electron beam current, while the pedestal position 
    remains stable. The $4\!:\!1$ area ratio between the beam signal peak and the pedestal of 
    each open histogram corresponds to the beam extraction cycle of $\;4\second\!:\!1\second$.
  }
\end{figure}

The filled grey histogram in Figure~\ref{fig:M4-Signal-DiffBC}(a) depicts the QDC response 
with no bias voltage applied to the photodetector (pedestal) and without ELSA operation, 
while the open histograms show the QDC response for a bias voltage of $400\:\volt$ applied 
to the photodetector (dark current). 
The dark (light) colour corresponds to the case without (with) beam circulating in ELSA. 
All three histograms are normalised to the same number of entries.  Both, the photodetector 
dark current and the accelerator operation, lead to a slight broadening of the pedestal peak, 
but its position remains stable. 
This is illustrated further by Figure~\ref{fig:M4-Signal-DiffBC}(b) which shows again a 
dark current signal (filled grey histogram) recorded while beam was circulating in ELSA, 
and, in addition, Cherenkov signals for three different extraction currents (open coloured 
histograms). Besides the beam signals, each open histogram features a small peak coinciding 
with the pedestal position because the data taking continued during the $1.1\:\second$ of 
filling and acceleration.  This provides the opportunity to monitor the pedestal stability 
continuously during beam operation. 
The relative areas of beam signal and pedestal peaks reflect the $4\!:\!1$ ratio defined 
by the $5.1\:\second$-periodic cycle of extraction and refill/acceleration times. 
Longterm monitoring showed that the pedestal position remained stable within $1$ QDC count,
which fulfills the ILC requirements.
All following Figures show pedestal-subtracted signals.

\subsection{Online alignment and channel response function} \label{sub:tb-align}
\begin{figure}[!h]
  \begin{picture}(16.0, 7.8)
    \put( 4.00, 0.00)  {\epsfig{file=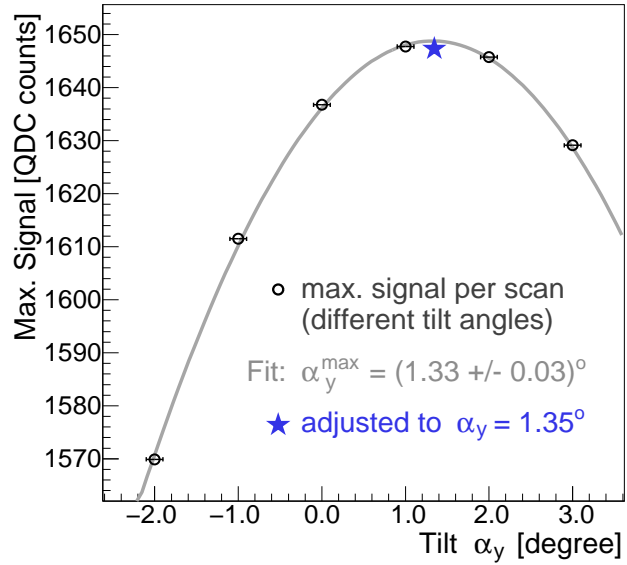, clip= , width=0.52\linewidth}}
  \end{picture}
  \caption{\label{fig:M4-TiltScan} \it 
    Detector alignment with the $2\!\times\!2$ MAPM (bias voltage $860\:\volt$): 
    The tilt in the ($x,z$)-plane is determined from $x$-scans for six different tilt angles. 
    An additional measurement for the adjusted tilt of $\alpha_y = 1.35^{\circ}$ 
    is also displayed.
  }
\end{figure}

The alignment of the detector with respect to the electron beam line was obtained from beam data.
By moving the detector stage, the incident beam position on the entrance window was scanned 
in horizontal ($x$) and vertical ($y$) directions. The adjustment procedure requires one 
vertical scan for each detector channel and a series of horizontal scans across both channels 
for different tilt angles $\alpha_y$, as shown in Figure~\ref{fig:M4-TiltScan}.

When the detector is tilted, the electrons will not have the full channel length to produce 
Cherenkov light. The maximal signal for any given $x$ position of the beam will be smaller 
than for a perfectly aligned detector~\footnote{Due to the channel geometry, 
  this holds for all tilt angles larger than $0.027^{\circ}$.}.
For each tilt angle, the beam $x$ position resulting in the highest signal is determined 
as displayed in Figure~\ref{fig:M4-TiltScan}. With this procedure, the best alignment of the detector with respect 
to the beam line was obtained for a tilt angle of $\alpha_y = 1.33^{\circ}$ with a very small statistical uncertainty of $0.03^{\circ}$. 
Due to the step size of the rotation mechanism, this value was approximated to
$\alpha^0_y = 1.35^{\circ}$ for all following measurements.

\begin{figure}[!h]
  \begin{picture}(16.0, 8.0)
    \put(-0.05, 0.00)  {\epsfig{file=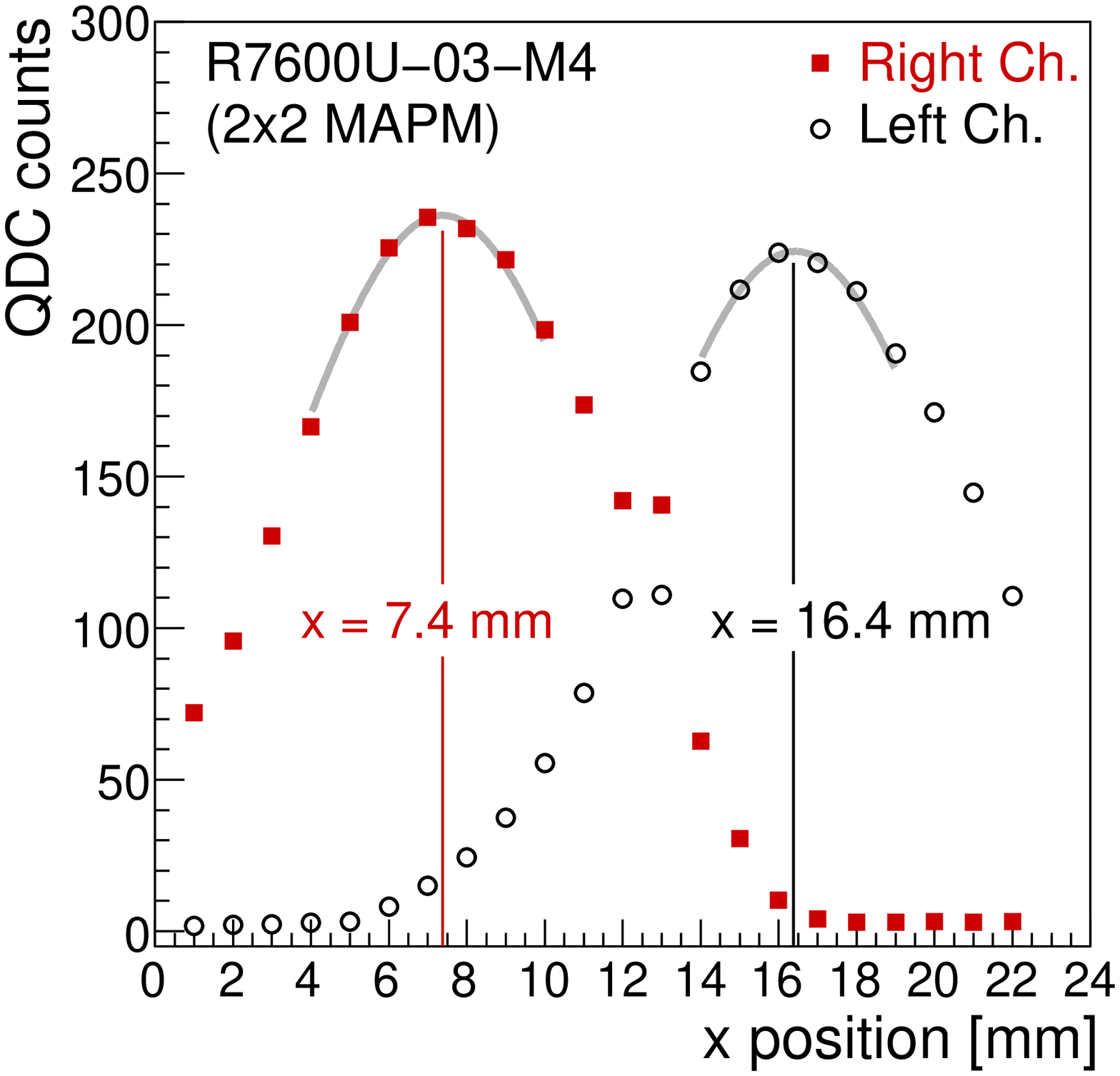,    clip= , width=0.50\linewidth}}
    \put( 8.10, 0.00)  {\epsfig{file=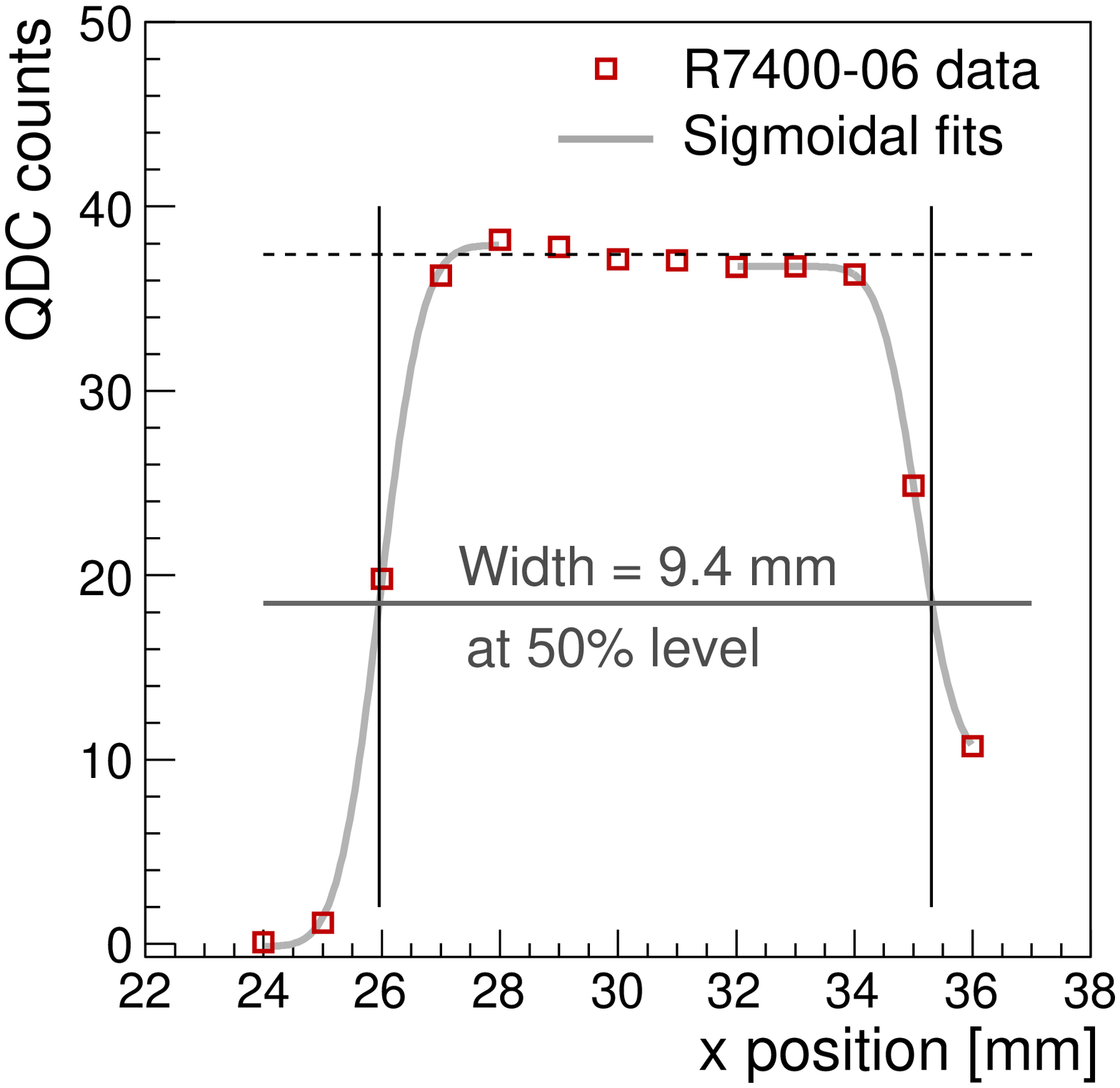, clip= , width=0.50\linewidth}}
    \put( 0.00, 0.00)  {(a)}
    \put( 8.35, 0.00)  {(b)}
  \end{picture}
  \caption{\label{fig:M4-R7400-Xscans} \it 
    Results from $x$-scans for two different types of photomultipliers: \newline 
    (a) the $2\!\times\!2$ MAPM (R7600U-03-M4, bias voltage $860\:\volt$), 
    (b) the SAPM (R7400U-06, bias voltage $300\:\volt$). 
    The absolute $x$-values correspond to different stage zero-positions; 
    only the relative values are relevant.
  }
\end{figure}

Measurements of the detector response as function of the horizontal beam entry position were performed with the single-anode photomultiplier R7400U-06 and with the $2\!\times\!2$ multi-anode photomultiplier (R7600U-03-M4) positioned on the detector channels as illustrated in Figure~\ref{fig:MountingPos}(c).
Figure~\ref{fig:M4-R7400-Xscans}(a) shows for the latter case the results of an $x$-scan across both detector channels.  
Two Gaussian fits indicate the respective channel centres to be at 
$x_{\rm right} =  (7.4 \pm 0.1)\:\milli\meter$ and 
$x_{\rm left}  = (16.4 \pm 0.1)\:\milli\meter$, leading to a distance 
of  $\Delta x   = (9.0 \pm 0.2)\:\milli\meter$. 
This agrees with the nominal distance between the channel centres of 
$\Delta x_{nom} =  8.8\:\milli\meter$, given by the width of one channel and the inter-channel wall. 

Figure~\ref{fig:M4-R7400-Xscans}(b) shows $x$-scan data for the single-anode photomultiplier, where a broad plateau is observed. The width of the signal region is determined 
from two sigmoidal fits to the edges of the plateau.  At 50\% of the plateau height, this width 
is found to be  $w = (9.4 \pm 0.3)\:\milli\meter$, where the error is dominated by the table 
position accuracy. This value is significantly larger than expected from the physical channel width 
and from Monte--Carlo simulations. Understanding the channel response function to the precision level required for the ILC will need further data with more stable beam conditions than available from the 2009 campaign.

\begin{figure}[!h]
  \begin{picture}(16.0, 7.5)
    \put( 0.05, 0.60)  {\epsfig{file=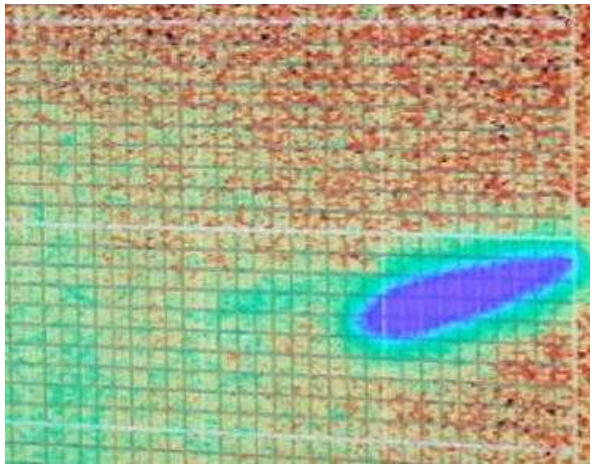, clip= , width=0.48\linewidth}}
    \put( 8.30, 0.60)  {\epsfig{file=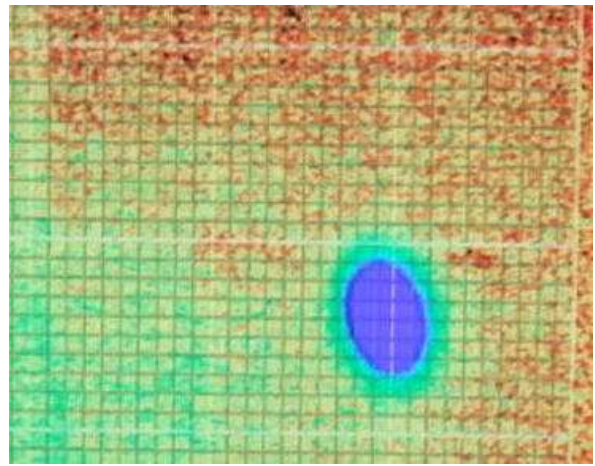,     clip= , width=0.48\linewidth}}
    \put( 0.10, 0.00)  {(a) larger, elongated beam spot}
    \put( 8.40, 0.00)  {(b) smaller, nearly round beam spot}
  \end{picture}
  \caption{\label{fig:BeamSpot} \it 
    Two beam spot shapes observed at ELSA when 
    (a) the data in Fig.$\;$\ref{fig:M4-R7400-Xscans}(a) and
    (b) the data in Fig.$\;$\ref{fig:M4-R7400-Xscans}(b) were recorded. 
      The dimensions of the spots are approximately 
      $4  \!\times\!2\:\milli\meter$ for the elongated shape in (a) and
      $1.5\!\times\!2\:\milli\meter$ for the rounder shape in (b).
  }
\end{figure}

The impact of the beam conditions becomes evident by comparing the observed channel reponse functions from~\ref{fig:M4-R7400-Xscans} with the corresponding photographs of a fluorescent screen placed on the detector entrance window: Figure~\ref{fig:BeamSpot}(a) shows a very elongated beam spot observed at the time the MAPM data were recorded (Fig.~\ref{fig:M4-R7400-Xscans}(a)), which explains the absence of any plateau in the detector response. The significantly smaller and nearly round beam spot shown in Figure~\ref{fig:BeamSpot}(b) was achieved during data taking with the SAPM (Fig.~\ref{fig:M4-R7400-Xscans}(b)).

\subsection{Alignment via spatial asymmetries}  \label{sub:tb-asym}
\begin{figure}[!h]
  \begin{picture}(6.4, 5.0)
    \put( 4.80, 0.00)  {\epsfig{file=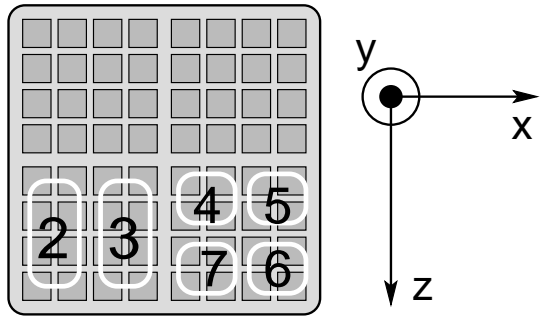, clip= , width=0.45\linewidth}}
  \end{picture}
  \caption{\label{fig:M64-ac6}  \it 
    Readout configuration for the $8\!\times\!8$ MAPM: The anode pads are 
    depicted as grey squares; the readout channels as numbered white rectangles.
  }
\end{figure}

The anode of the 8$\times$8 multi-anode photomultiplier (R7600-00-M64) is finely segmented 
with $16$~anode pads covering a single Cherenkov channel, thus offering spatial resolution 
within a detector channel. Since two QDC channels were broken, only six channels were 
available to realise the readout configuration illustrated in Figure~\ref{fig:M64-ac6}.
The numbers indicate the QDC channel utilised to read out the sum signal of either 
four or eight anode pads of the photodetector.

\begin{figure}[!h]
  \begin{picture}(16.0, 7.8)
    \put(-0.05, 0.00)  {\epsfig{file=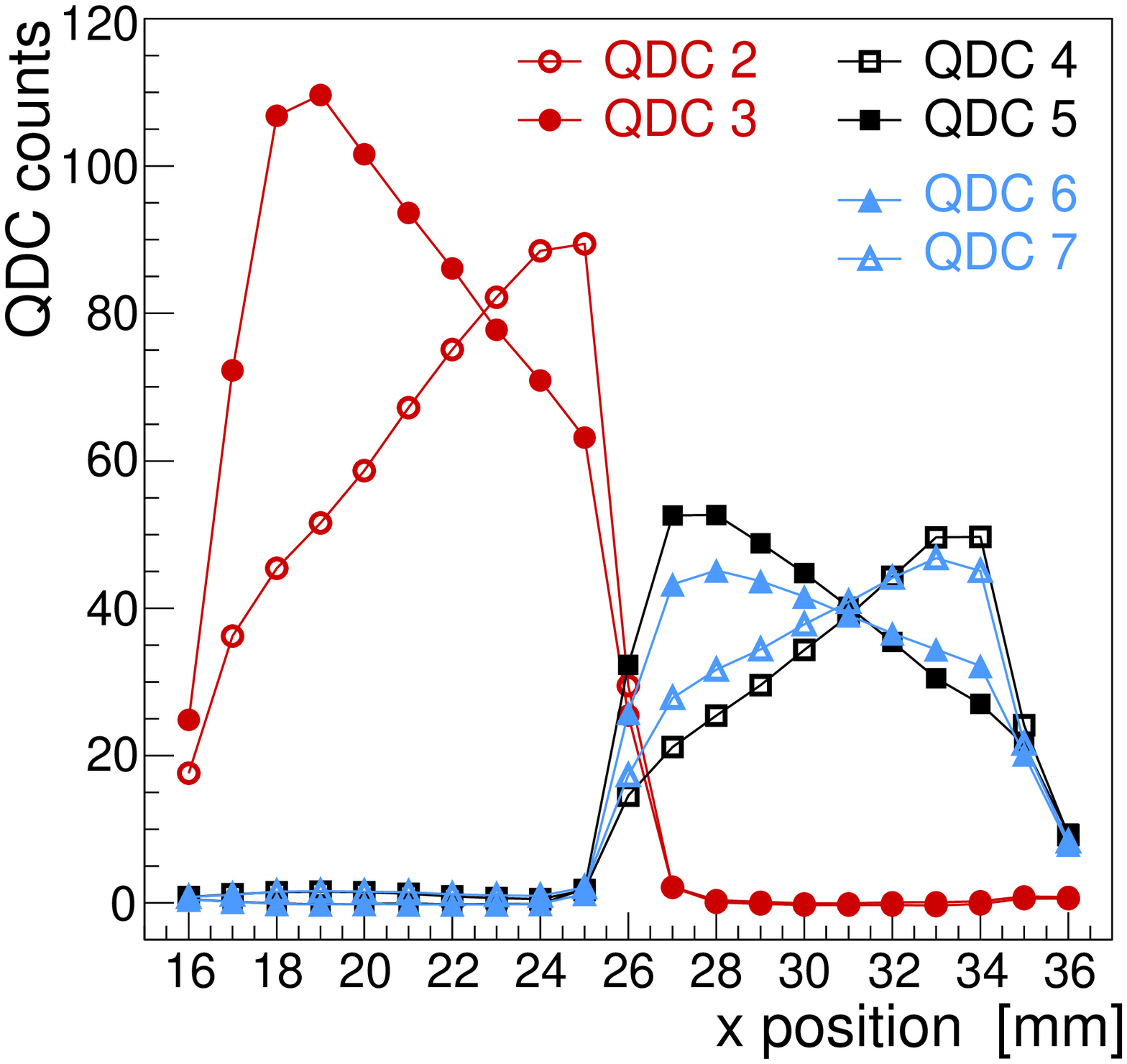, clip= , width=0.50\linewidth}}
    \put( 8.10, 0.00)  {\epsfig{file=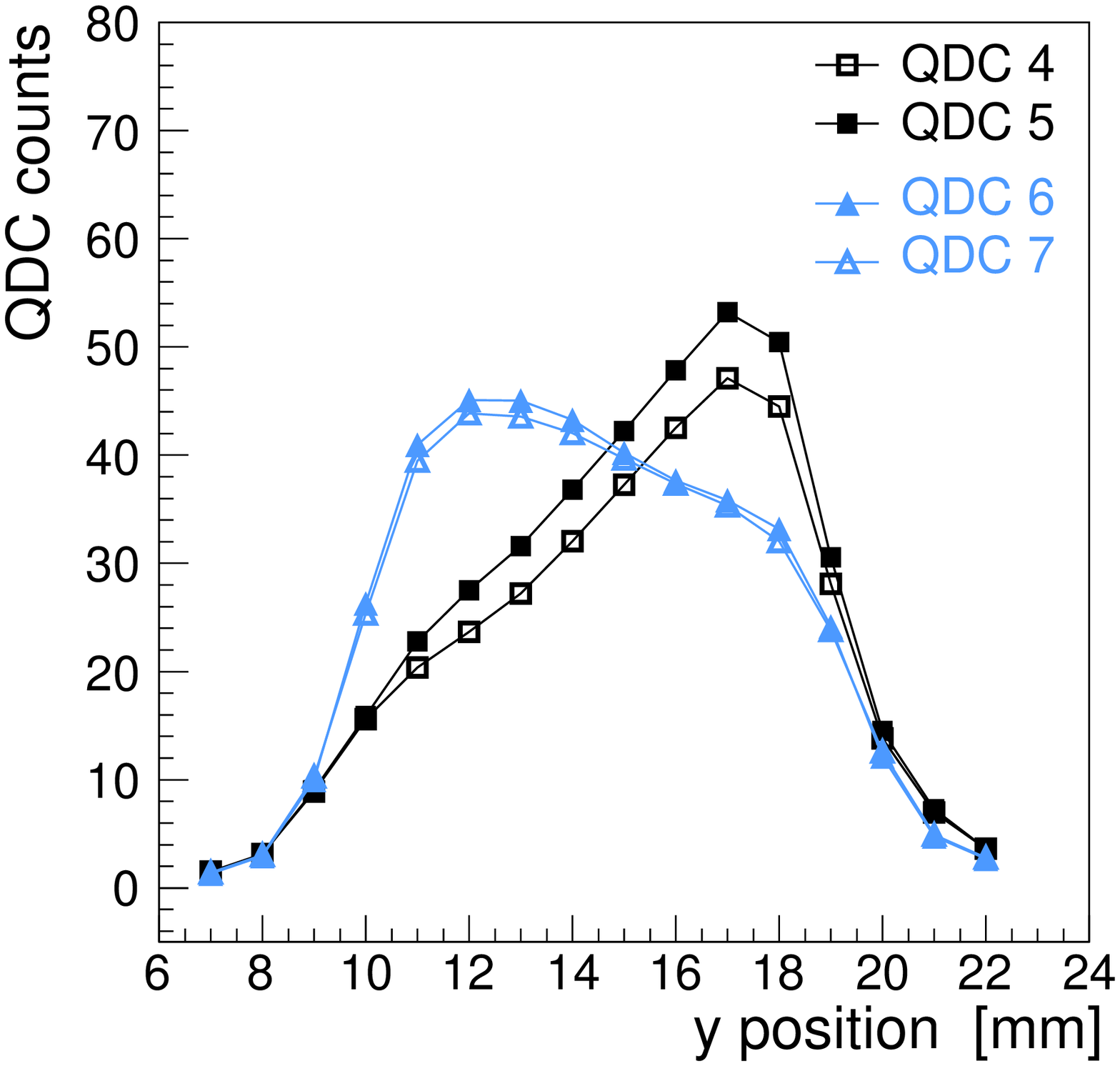, clip= , width=0.50\linewidth}}
    \put( 0.00, 0.00)  {(a)}
    \put( 8.35, 0.00)  {(b)}
  \end{picture}
  \caption{\label{fig:M64-XYscan-PedSub} \it 
    Beam postion scan data recorded with the $8\!\times\!8$ MAPM (bias voltage $500\:\volt$): 
    (a) $x$-scan across both channels and 
    (b) $y$-scan on the left channel.
  }
\end{figure}

Figure~\ref{fig:M64-XYscan-PedSub} shows the results of 
(a) an $x$-scan across both detector channels and 
(b) the corresponding $y$-scan across the left channel. 
As expected, the signals in QDC channels~$2$ and~$3$ are about twice as large as in 
the other channels since eight instead of only four anode pads are grouped together. 
The asymmetric response reflects the incident beam position.  For each QDC channel, 
the largest signal is observed when the beam enters on the opposite side of the 
detector channel. This confirms the prediction of one glancing angle reflection for 
most of the photons obtained from MC simulations 
(c.f.\ Section~\ref{sub:sim-lightdist} and~\ref{sub:sim-asym}). 

\begin{figure}[!h]
  \begin{picture}(16.0, 7.8)
    \put(-0.05, 0.00)  {\epsfig{file=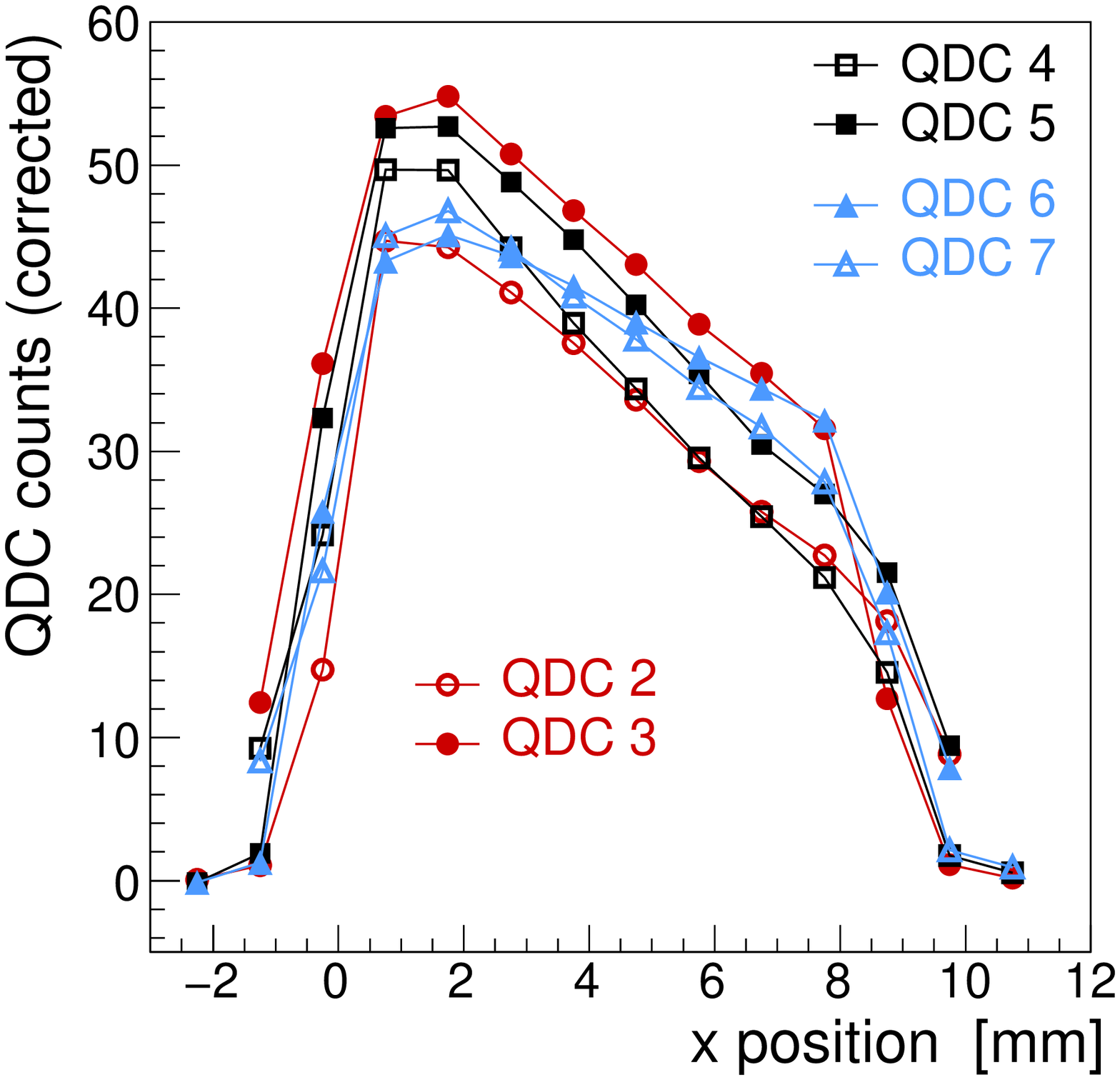, clip= , width=0.50\linewidth}}
    \put( 8.10, 0.00)  {\epsfig{file=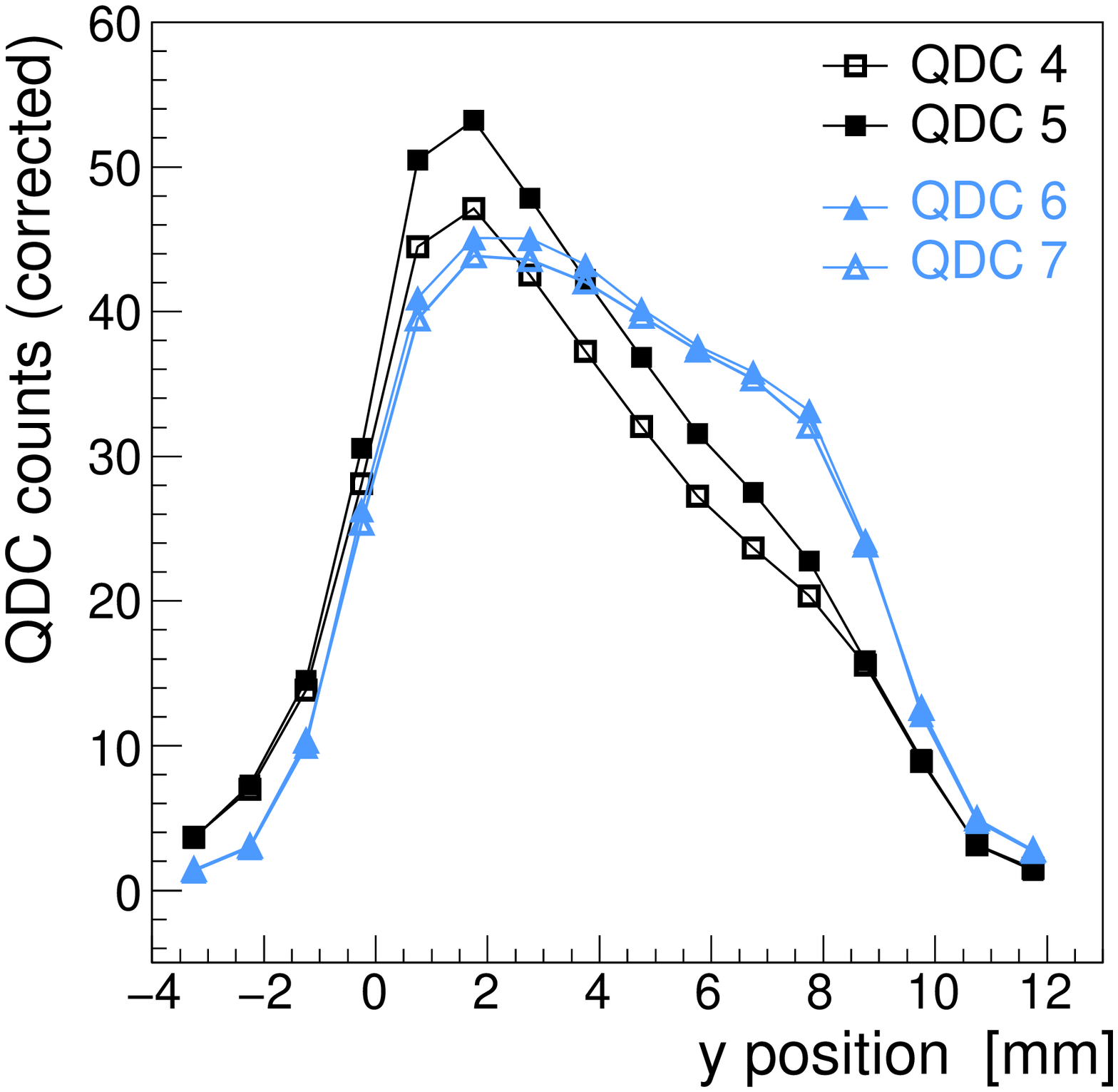, clip= , width=0.50\linewidth}}
    \put( 0.00, 0.00)  {(a)}
    \put( 8.35, 0.00)  {(b)}
  \end{picture}
  \caption{\label{fig:M64-XYscan-refined} \it 
    Different visualisation of the position scan data presented in Figure~\ref{fig:M64-XYscan-PedSub} 
    with the emphasis on shape and amplitude differences:  (a) $x$-scan data and (b) $y$-scan data.
  }
\end{figure}

For a more detailed comparison of the responses of the different anode segments, 
the same data are displayed again in Figure~\ref{fig:M64-XYscan-refined}, scaled and 
mirrored to correct for the two above effects. Possible reasons for the remaining 
shape and amplitude differences comprise gain variations between the pads and 
residual detector misalignment. 

The vertical beam scan data have been used to calibrate the relative gain variations between the
different groups of anodes. After applying these calibrations to the horizontal beam scans, two $x$-asymmetries, $A_x^{lower}$ and $A_x^{upper}$, are calculated from the four anode groups QDC$\;4$ to QDC$\;7$ (c.f. Figure~\ref{fig:M64-ac6}):  

\small
\begin{align*}
  \mbox{\large$A_x^{lower}$}  = \frac{\mathrm{QDC}\,5\,-\,\mathrm{QDC}\,4} {\mathrm{QDC}\,5\,+\,\mathrm{QDC}\,4} \hspace*{15mm} & \hspace*{12mm} 
   \mbox{\large$A_x^{upper}$}  = \frac{\mathrm{QDC}\,6\,-\,\mathrm{QDC}\,7} {\mathrm{QDC}\,6\,+\,\mathrm{QDC}\,7} \hspace*{15mm}\\[4mm]
\end{align*}
\normalsize

\begin{figure}[!h]
  \begin{picture}(16.0, 8.0)
    \put(-0.05, 0.00)  {\epsfig{file=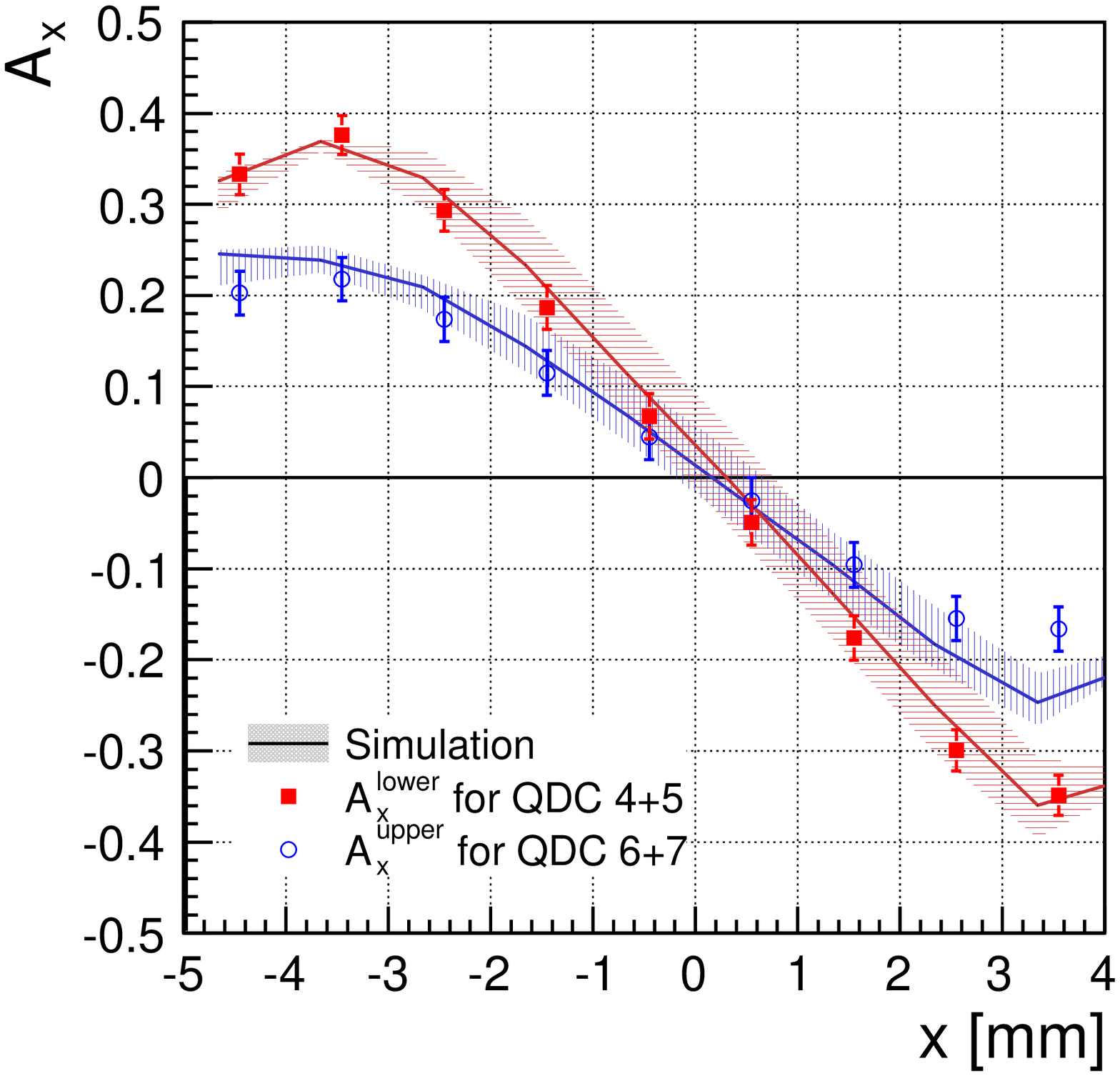, bb=20 30 550 513, clip= , width=0.50\linewidth}}
    \put( 8.15, 0.00)  {\epsfig{file=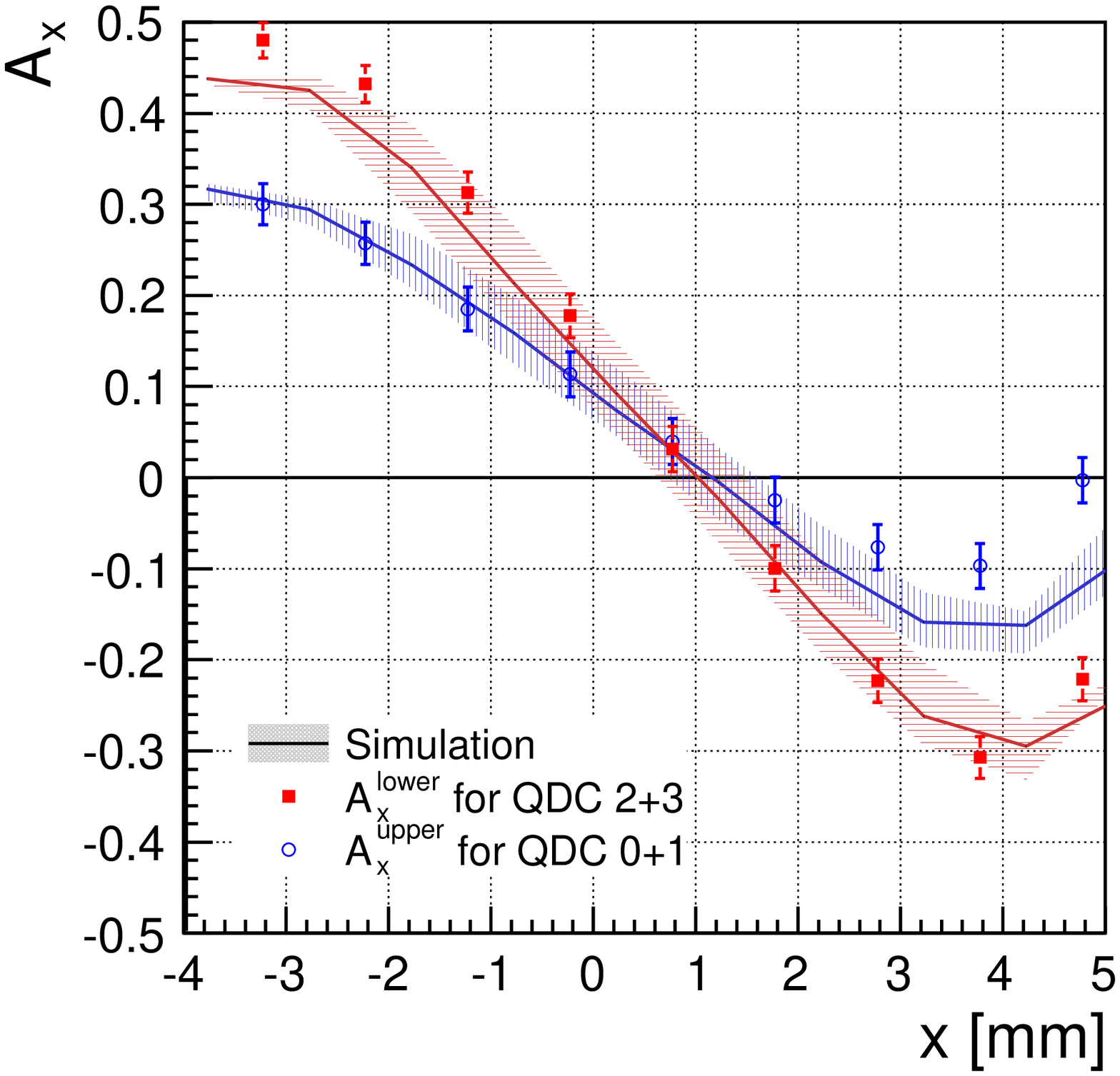, bb=20 30 550 525, clip= , width=0.50\linewidth}}
    \put( 0.00, 0.00)  {(a)}
    \put( 8.35, 0.00)  {(b)}
  \end{picture}
  \caption{\label{fig:AsymXZ-data} \it 
    Asymmetries calculated from the position scan data sets 
    recorded with the $8\!\times\!8$ MAPM.  
    (a) $A_x$ for QDC-pairings~$4\!+\!5$ and~$6\!+\!7$ on the left detector channel from the same data as used in Fig~\ref{fig:M64-XYscan-refined} 
    (b) $A_x$ on the right detector channel from a subsequent run.
    In addition simulated asymmetries for $\alpha_x = 0.2^{\circ}$ and $\alpha_y =   
    -0.2^{\circ}$ are shown for both channels. The error bands correspond to a 
    variation of $\alpha_y$ by $\pm 0.1^{\circ}$. 
  }
\end{figure}

The resulting asymmetries are displayed in Figure~\ref{fig:AsymXZ-data}(a), together with the 
expectation from simulation assuming residual tilts of $\alpha_x = 0.2^{\circ}$ and $\alpha_y = -0.2^{\circ}$ and a reflectivity ratio of $r = 85\%$. The error bars on the data points correspond to the remaining gain differences between the anode pads. Uncertainties which are in common between the different pads cancel out within the asymmetries. The error bands on the simulated curves are obtained by varying $\alpha_y$ by $\pm 0.1^{\circ}$ in the simulation. A similar variation of $\alpha_x$ has no visible impact on $A_x$. Figure~\ref{fig:AsymXZ-data}(b) shows $A_x^{lower}$ and $A_x^{upper}$ for the other detector channel, read-out by the neighboring quadrant of the $8\!\times\!8$ MAPM\footnote{Due to the two broken QDC channels, these data could not be taken simultaneously, but stem from a subsequent run after changing the readout combination of anode segments at the QDC.}, compared to the simulation using exactly the same parameters as before. The obtained precision of $\pm 0.1^{\circ}$ is very close to the requirements for the ILC, which shows that in principle the light distribution inside a channel can be understood to sufficient precision. But due to the extremely limited number of datasets with good beam conditions (c.f. Section~\ref{sub:tb-align}) no further independent beam scan is available from the 2009 data-taking period at ELSA. Therefore further data-taking will be needed for conclusive statements on the use of multi-anode photodetectors in ILC polarimeters.

\section{Conclusions}  \label{sec:conclusion}
At a future $e^+e^-$ linear collider, Compton polarimeters will be employed to measure 
the beam polarisation to a precision of $\delta{\mathcal P}/{\mathcal P}=0.25$\%, using 
Cherenkov detectors to register the scattered Compton electrons.

A compact two channel prototype detector has been designed and constructed such that it will allow nearly all aspects of the final detector to be studied. 
In particular, it has been designed for easy exchange of the photodetectors and 
the calibration light source, but also the inter-channel wall could be exchanged in order to test different materials for a final detector. 

The prototype has been operated successfully in a first testbeam campaign using four different photodetectors. The dynamic range of the detector and the pedestal stability fulfill the ILC requirements. The measurements have been compared to a detailed simulation of the prototype and 
several alignment methods have been tested.

In particular, a method to extract intra-channel beam position information has been developed, which could possibly allow to calibrate the Compton edge position without need for dedicated beam-time. Furthermore, the detector response has been studied as a function of the beam position. 
This will lead to a determination of each channel's response function which is important in order to control systematic effects on the final polarisation measurements.

In the future it is planned to use this prototype to compare different photodetectors and wall materials as well as to establish a calibration to the permille level as required for the ILC.

\section*{Acknowledgements}  \label{sec:acknowledge}
We thank C.~Hagner and R.~Klanner of the Institut f\"ur Experimentalphysik of the 
University of Hamburg for their support, and especially the design and technical 
construction team, B.~Frensche and J.~Pelz, as well as the head of the mechanical 
workshop, S.~Fleig, and his entire team for their competent work. 

The authors are grateful to W.~Hillert, F.~Frommberger and the entire ELSA team 
for their support during the testbeam period, for realising special beam requests and 
for many helpful discussions concerning a multitude of different ELSA technicalities.
Further thanks go to K.~Desch, J.~Kaminski, D.~Elsner, and many others for help and 
support with the planning and setup before, during, and after the two weeks in Bonn. 

The authors acknowledge the financial support of 
the Deutsche Forschungsgemeinschaft in the DFG project Li~1560/1-1 
as well as the complementary support by 
the Initative and Networking Fund of the Helmholtz Association, 
contract HA-101 (``Physics at the Terascale'').

\begin{footnotesize}


\end{footnotesize}

\clearpage

\begin{table}[htb]
  \renewcommand{\arraystretch}{1.10}
  \begin{tabular*}{0.47\textwidth }{@{\extracolsep{\fill}}ccc}
    \hline\hline 
    wavelength           & $R_{\rm diam}$  &  $R_{\rm roll}$ \\
    \hline\hline 
    $160\:\nano\meter$   &    $74\:\%$     &     $11\:\%$    \\
    $180\:\nano\meter$   &    $77\:\%$     &     $18\:\%$    \\
    $200\:\nano\meter$   &    $81\:\%$     &     $27\:\%$    \\
    $220\:\nano\meter$   &    $84\:\%$     &     $30\:\%$    \\
    $240\:\nano\meter$   &    $86\:\%$     &     $37\:\%$    \\
    $500\:\nano\meter$   &    $85\:\%$     &     $40\:\%$    \\
    $520\:\nano\meter$   &    $84\:\%$     &     $39\:\%$    \\
    $650\:\nano\meter$   &    $83\:\%$     &     $40\:\%$    \\
    $900\:\nano\meter$   &    $82\:\%$     &     $39\:\%$    \\
    \hline\hline 
  \end{tabular*}
  \caption[Surface reflectivity measurements]{\it  
    The reflectivities of 
    diamond-milled quality aluminium $R_{\rm diam}$ 
    and  of rolled quality aluminium $R_{\rm roll}$ as determined with the 
    \perkinelmer{} spectrometer and implemented in the \geant{} simulation. For the rolled aluminium, only the wavelength dependency is transfered to the simulation, while the absolute normalisation is adjusted to prototype data.}
  \label{tab:reflect-meas}
\end{table} 
\begin{table}[htb]
  \renewcommand{\arraystretch}{1.10}
  \begin{tabular*}{\textwidth }{@{\extracolsep{\fill}}l c cc rr}
    \hline\hline 
    $\;$photodetector            &        sensitive area          &     wavelength  &     typical      &  response              & anode \\
    $\;\;\;$types                &      in [$\milli\meter^2$]     &     range [nm]  &      gain        & time$\;\;\;$           &  pads \\
    \hline\hline 
    \multicolumn{6}{c}{} \\[-4.8mm]
    $\;$R7600U-03-M4$^{\;(a)}$   &      $18.0\times18.0$          &      185 - 600  &  $1.8\cdot10^6$  &  $11.0\:\nano\second$  &    4  \\
    $\;$R7600-00-M64$^{\;(a)}$   &      $18.1\times18.1$          &      300 - 600  &  $0.3\cdot10^6$  &  $13.4\:\nano\second$  &   64  \\
    \multicolumn{6}{c}{} \\[-4.8mm] \hline 
    \multicolumn{6}{c}{} \\[-4.2mm] 
    $\;$R7400U-06(03)$^{\;(a)}$  &  $\diameter= 8\:\milli\meter$  & 160(185) - 600  &  $0.7\cdot10^6$  &  $ 6.5\:\nano\second$  &    1  \\
    $\;$XP1911/UV$^{\;(b)}$      &  $\diameter=15\:\milli\meter$  &      200 - 600  &  $0.9\cdot10^6$  &  $28.0\:\nano\second$  &    1  \\
    \hline\hline 
    \multicolumn{6}{l}{\scriptsize Photodetector from:\quad (a) \hamamatsu,\quad (b) \photonis.}
  \end{tabular*}
  \caption{\it  Key characteristics of the four different photomultipliers 
    from \hamamatsu{} and \photonis~\cite{bib:Hamamatsu, bib:Photonis}. 
    The two MAPMs (R7600U-03-M4 and R7600-00-M64) have a quadratic cross-section of 
    similar size, but differ in the number of anodes and in their wavelength range. 
    The two SAPMs (R7400U-03 and R7400U-06) differ in the size of 
    their sensitive areas and slightly in wavelength range.}
  \label{tab:PM-types}
\end{table}
\clearpage

\clearpage

\clearpage

\clearpage





\clearpage

\clearpage

\clearpage


\end{document}